\documentclass[a4paper,11pt]{article}

\usepackage{amsfonts}
\usepackage{amsmath}
\usepackage{amssymb}
\usepackage{algorithm}
\usepackage{authblk}
\usepackage{amsfonts,amsmath,amssymb,amsthm}
\usepackage{paralist}
\usepackage{url}
\usepackage{hyperref}
\usepackage{graphicx}
\usepackage{microtype}
\usepackage{subcaption}
\usepackage{url}
\usepackage{hyperref}
\usepackage{tikz}
\usepackage{paralist}
\usepackage{xparse}
\usetikzlibrary{positioning}
\usetikzlibrary{decorations.pathreplacing}
\definecolor{myyellow}{HTML}{FFFF00}
\definecolor{myred}{HTML}{FF4500}
\definecolor{mygreen}{HTML}{32CD32}
\definecolor{myblue}{HTML}{1E90FF}

\makeatletter

\pgfdeclareshape{rectangle with diagonal fill}
{
	\inheritsavedanchors[from=rectangle]
	\inheritanchorborder[from=rectangle]
	\inheritanchor[from=rectangle]{north}
	\inheritanchor[from=rectangle]{north west}
	\inheritanchor[from=rectangle]{north east}
	\inheritanchor[from=rectangle]{center}
	\inheritanchor[from=rectangle]{west}
	\inheritanchor[from=rectangle]{east}
	\inheritanchor[from=rectangle]{mid}
	\inheritanchor[from=rectangle]{mid west}
	\inheritanchor[from=rectangle]{mid east}
	\inheritanchor[from=rectangle]{base}
	\inheritanchor[from=rectangle]{base west}
	\inheritanchor[from=rectangle]{base east}
	\inheritanchor[from=rectangle]{south}
	\inheritanchor[from=rectangle]{south west}
	\inheritanchor[from=rectangle]{south east}
	
	\inheritbackgroundpath[from=rectangle]
	\inheritbeforebackgroundpath[from=rectangle]
	\inheritbehindforegroundpath[from=rectangle]
	\inheritforegroundpath[from=rectangle]
	\inheritbeforeforegroundpath[from=rectangle]
	
	\behindbackgroundpath{%
		\pgfextractx{\pgf@xa}{\southwest}%
		\pgfextracty{\pgf@ya}{\southwest}%
		\pgfextractx{\pgf@xb}{\northeast}%
		\pgfextracty{\pgf@yb}{\northeast}%
		\ifpgf@diagonal@lefttoright
		\def\pgf@diagonal@point@a{\pgfpoint{\pgf@xa}{\pgf@yb}}%
		\def\pgf@diagonal@point@b{\pgfpoint{\pgf@xb}{\pgf@ya}}%
		\else
		\def\pgf@diagonal@point@a{\southwest}%
		\def\pgf@diagonal@point@b{\northeast}%
		\fi
		\pgfpathmoveto{\pgf@diagonal@point@a}%
		\pgfpathlineto{\northeast}%
		\pgfpathlineto{\pgfpoint{\pgf@xb}{\pgf@ya}}%
		\pgfpathclose
		\ifpgf@diagonal@lefttoright
		\color{\pgf@diagonal@top@color}%
		\else
		\color{\pgf@diagonal@bottom@color}%
		\fi
		\pgfusepath{fill}%
		\pgfpathmoveto{\pgfpoint{\pgf@xa}{\pgf@yb}}%
		\pgfpathlineto{\southwest}%
		\pgfpathlineto{\pgf@diagonal@point@b}%
		\pgfpathclose
		\ifpgf@diagonal@lefttoright
		\color{\pgf@diagonal@bottom@color}%
		\else
		\color{\pgf@diagonal@top@color}%
		\fi
		\pgfusepath{fill}%
	}
}

\newif\ifpgf@diagonal@lefttoright
\def\pgf@diagonal@top@color{white}
\def\pgf@diagonal@bottom@color{gray!30}

\def\pgfsetdiagonaltopcolor#1{\def\pgf@diagonal@top@color{#1}}%
\def\pgfsetdiagonalbottomcolor#1{\def\pgf@diagonal@bottom@color{#1}}%
\def\pgfsetdiagonallefttoright{\pgf@diagonal@lefttorighttrue}%
\def\pgfsetdiagonalrighttoleft{\pgf@diagonal@lefttorightfalse}%

\tikzoption{diagonal top color}{\pgfsetdiagonaltopcolor{#1}}
\tikzoption{diagonal bottom color}{\pgfsetdiagonalbottomcolor{#1}}
\tikzoption{diagonal from left to right}[]{\pgfsetdiagonallefttoright}
\tikzoption{diagonal from right to left}[]{\pgfsetdiagonalrighttoleft}

\makeatother

\tikzset{
    mybrace/.style={decorate,decoration={brace,aspect=#1}}
}
\usetikzlibrary{matrix,backgrounds}
\pgfdeclarelayer{myback}
\pgfsetlayers{myback,background,main}

\tikzset{mycolor/.style = {line width=1bp,color=#1}}%
\tikzset{myfillcolor/.style = {draw,fill=#1}}%

\NewDocumentCommand{\highlight}{O{blue!40} m m}{%
\draw[mycolor=#1] (#2.north west)rectangle (#3.south east);
}

\NewDocumentCommand{\fhighlight}{O{blue!40} m m}{%
\draw[myfillcolor=#1] (#2.north west)rectangle (#3.south east);
}

\newcommand{\N}{\mathbb{N}}

\newcommand{\F}{\mathbb{F}}

\newtheorem{theorem}{Theorem}%
\newtheorem{lemma}[theorem]{Lemma}%
\newtheorem{definition}{Definition}%
\newtheorem{problem}{Problem}%

\tikzset{
    mybrace/.style={decorate,decoration={brace,aspect=#1}}
}

\providecommand{\keywords}[1]{\textbf{\textit{Keywords }} #1}

\begin{document}

\title{Enumeration of Maximal Cycles Generated by Orthogonal Cellular Automata}

\author[1]{Luca Mariot}
	
\affil[1]{{\small Digital Security Group, Radboud University, PO Box 9010, Nijmegen, The Netherlands} 
	
	{\small \texttt{luca.mariot@ru.nl}}}

\maketitle

\begin{abstract}
Cellular Automata (CA) are an interesting computational model for designing Pseudorandom Number Generators (PRNG), due to the complex dynamical behavior they can exhibit depending on the underlying local rule. Most of the CA-based PRNGs proposed in the literature, however, suffer from poor diffusion since a change in a single cell can propagate only within its neighborhood during a single time step. This might pose a problem especially when such PRNGs are used for cryptographic purposes. In this paper, we consider an alternative approach to generate pseudorandom sequences through \emph{orthogonal CA} (OCA), which guarantees a better amount of diffusion. After defining the related PRNG, we perform an empirical investigation of the maximal cycles in OCA pairs up to diameter $d=8$. Next, we focus on OCA induced by linear rules, giving a characterization of their cycle structure based on the rational canonical form of the associated Sylvester matrix. Finally, we devise an algorithm to enumerate all linear OCA pairs characterized by a single maximal cycle, and apply it up to diameter $d=16$ and $d=13$ for OCA respectively over the binary and ternary alphabets.
\end{abstract}


\section{Introduction}
\label{sec:intro}
\begin{figure}[b]
    \centering
    \includegraphics[height=0.18\textheight]{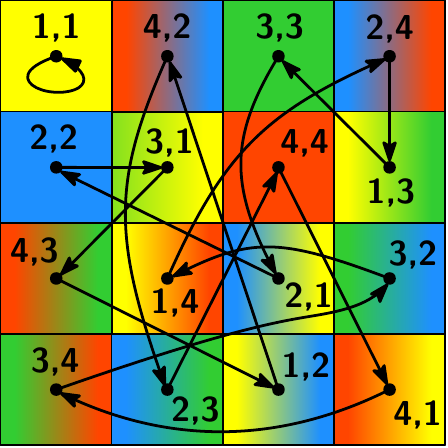}
    \caption{Example of square game. Choosing any initial cell other than the one at the top left corner gives a maximum score of 15.}
    \label{fig:ex-game}
\end{figure}

Consider the following game: we are given a $N\times N$ square, where each cell is labelled by a pair of numbers $(i,j)$ with $i,j \in \{1,\cdots, N\} = [N]$. Moreover, we assume that each of the $N^2$ pairs in the Cartesian product $[N]\times[N]$ occurs exactly once as a label in the square. Our only move is to choose an initial cell; after that, we read the corresponding label $(i,j)$, and use it as the new row and column coordinates of the cell where to jump next. The process is then iterated until we jump back to the initial cell, which is granted by the assumption that the cells' labeling is a permutation of $[N] \times [N]$. The goal of the game is to achieve the highest score, defined as the number of distinct cells visited before returning to the initial one. Figure~\ref{fig:ex-game} depicts an example of $4 \times 4$ square where choosing any initial cell except the top left one always yields the highest score, which is 15 in this case.

Given the rules above, ``winning'' the game depends on two factors: 1) the cycle structure of the permutation that defines the cells' labeling, and 2) the initial cell where we start from. Clearly, there is a trade-off between these two aspects: the more the cells' labeling permutation is composed of few cycles having a large length, the less the position of the initial cell matters to reach a high score. Figure~\ref{fig:ex-game} represents an extreme case, where the permutation is made only of a single large cycle of length $2^N - 1$ and a fixed point.

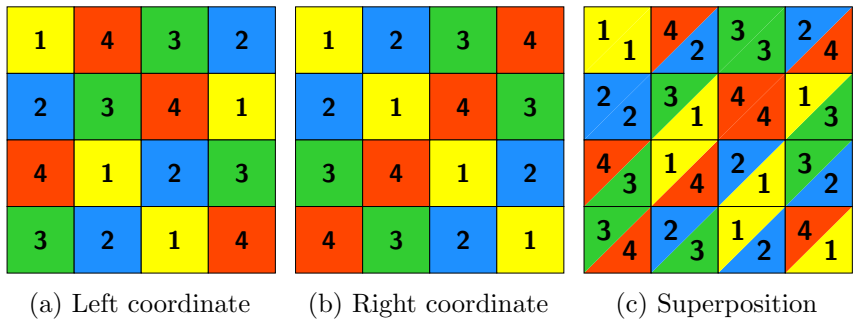
\begin{figure}[t]
\centering
\begin{subfigure}{.3\textwidth}
\centering
\resizebox{0.18\textheight}{!}{
\begin{tikzpicture}
[->,auto,node distance=1.5cm,
empt node/.style={font=\sffamily,inner sep=0pt,minimum size=0pt},
rect1 node/.style={rectangle,draw,fill=myyellow,font=\sffamily\bfseries,minimum size=1cm, inner
	sep=0pt, outer sep=0pt},
rect2 node/.style={rectangle,draw,fill=myblue,font=\sffamily\bfseries,minimum size=1cm, inner
	sep=0pt, outer sep=0pt},
rect3 node/.style={rectangle,draw,fill=mygreen,font=\sffamily\bfseries,minimum size=1cm, inner
	sep=0pt, outer sep=0pt},
rect4 node/.style={rectangle,draw,fill=myred,font=\sffamily\bfseries,minimum size=1cm, inner
	sep=0pt, outer sep=0pt}]

\node [rect1 node] (s11) {1};
\node [rect4 node] (s12) [right=0cm of s11] {4};
\node [rect3 node] (s13) [right=0cm of s12] {3};
\node [rect2 node] (s14) [right=0cm of s13] {2};

\node [rect2 node] (s21) [below=0cm of s11] {2};
\node [rect3 node] (s22) [right=0cm of s21] {3};
\node [rect4 node] (s23) [right=0cm of s22] {4};
\node [rect1 node] (s24) [right=0cm of s23] {1};

\node [rect4 node] (s31) [below=0cm of s21] {4};
\node [rect1 node] (s32) [right=0cm of s31] {1};
\node [rect2 node] (s33) [right=0cm of s32] {2};
\node [rect3 node] (s34) [right=0cm of s33] {3};

\node [rect3 node] (s41) [below=0cm of s31] {3};
\node [rect2 node] (s42) [right=0cm of s41] {2};
\node [rect1 node] (s43) [right=0cm of s42] {1};
\node [rect4 node] (s44) [right=0cm of s43] {4};

\end{tikzpicture}
}
\caption{Left coordinate}
\end{subfigure}%
\begin{subfigure}{.3\textwidth}
\centering
\resizebox{0.18\textheight}{!}{
\begin{tikzpicture}
[->,auto,node distance=1.5cm,
empt node/.style={font=\sffamily,inner sep=0pt,minimum size=0pt},
rect1 node/.style={rectangle,draw,fill=myyellow,font=\sffamily\bfseries,minimum size=1cm, inner
	sep=0pt, outer sep=0pt},
rect2 node/.style={rectangle,draw,fill=myblue,font=\sffamily\bfseries,minimum size=1cm, inner
	sep=0pt, outer sep=0pt},
rect3 node/.style={rectangle,draw,fill=mygreen,font=\sffamily\bfseries,minimum size=1cm, inner
	sep=0pt, outer sep=0pt},
rect4 node/.style={rectangle,draw,fill=myred,font=\sffamily\bfseries,minimum size=1cm, inner
	sep=0pt, outer sep=0pt}]

\node [rect1 node] (s11) {1};
\node [rect2 node] (s12) [right=0cm of s11] {2};
\node [rect3 node] (s13) [right=0cm of s12] {3};
\node [rect4 node] (s14) [right=0cm of s13] {4};

\node [rect2 node] (s21) [below=0cm of s11] {2};
\node [rect1 node] (s22) [right=0cm of s21] {1};
\node [rect4 node] (s23) [right=0cm of s22] {4};
\node [rect3 node] (s24) [right=0cm of s23] {3};

\node [rect3 node] (s31) [below=0cm of s21] {3};
\node [rect4 node] (s32) [right=0cm of s31] {4};
\node [rect1 node] (s33) [right=0cm of s32] {1};
\node [rect2 node] (s34) [right=0cm of s33] {2};

\node [rect4 node] (s41) [below=0cm of s31] {4};
\node [rect3 node] (s42) [right=0cm of s41] {3};
\node [rect2 node] (s43) [right=0cm of s42] {2};
\node [rect1 node] (s44) [right=0cm of s43] {1};

\end{tikzpicture}
}
\caption{Right coordinate}
\end{subfigure}%
\begin{subfigure}{.3\textwidth}
\centering
\resizebox{0.18\textheight}{!}{
\begin{tikzpicture}
[->,auto,node distance=1.5cm,
empt node/.style={font=\sffamily,inner sep=0pt,minimum size=0pt},
rect node/.style={rectangle,draw,font=\sffamily,minimum size=0.7cm, inner sep=0pt, outer sep=0pt}]

\node[rectangle with diagonal fill,
diagonal top color=myyellow,
diagonal bottom color=myyellow,
diagonal from right to left,
draw,font=\sffamily\bfseries,minimum size=1cm, inner sep=0pt, outer
sep=0pt] (s11) {$^{\textsf{\large 1}}$ $_{\textsf{\large 1}}$};
\node[rectangle with diagonal fill,
diagonal top color=myred,
diagonal bottom color=myblue,
diagonal from right to left,
draw,font=\sffamily\bfseries,minimum size=1cm, inner sep=0pt, outer
sep=0pt] (s12) [right=0cm of s11] {$^{\textsf{\large 4}}$ $_{\textsf{\large 2}}$};
\node[rectangle with diagonal fill,
diagonal top color=mygreen,
diagonal bottom color=mygreen,
diagonal from right to left,
draw,font=\sffamily\bfseries,minimum size=1cm, inner sep=0pt, outer
sep=0pt] (s13) [right=0cm of s12] {$^{\textsf{\large 3}}$ $_{\textsf{\large 3}}$};
\node[rectangle with diagonal fill,
diagonal top color=myblue,
diagonal bottom color=myred,
diagonal from right to left,
draw,font=\sffamily\bfseries,minimum size=1cm, inner sep=0pt, outer
sep=0pt] (s14) [right=0cm of s13] {$^{\textsf{\large 2}}$ $_{\textsf{\large 4}}$};

\node[rectangle with diagonal fill,
diagonal top color=myblue,
diagonal bottom color=myblue,
diagonal from right to left,
draw,font=\sffamily\bfseries,minimum size=1cm, inner sep=0pt, outer
sep=0pt] (s21) [below=0cm of s11] {$^{\textsf{\large 2}}$ $_{\textsf{\large 2}}$};
\node[rectangle with diagonal fill,
diagonal top color=mygreen,
diagonal bottom color=myyellow,
diagonal from right to left,
draw,font=\sffamily\bfseries,minimum size=1cm, inner sep=0pt, outer
sep=0pt] (s22) [right=0cm of s21] {$^{\textsf{\large 3}}$ $_{\textsf{\large 1}}$};
\node[rectangle with diagonal fill,
diagonal top color=myred,
diagonal bottom color=myred,
diagonal from right to left,
draw,font=\sffamily\bfseries,minimum size=1cm, inner sep=0pt, outer
sep=0pt] (s23) [right=0cm of s22] {$^{\textsf{\large 4}}$ $_{\textsf{\large 4}}$};
\node[rectangle with diagonal fill,
diagonal top color=myyellow,
diagonal bottom color=mygreen,
diagonal from right to left,
draw,font=\sffamily\bfseries,minimum size=1cm, inner sep=0pt, outer
sep=0pt] (s24) [right=0cm of s23] {$^{\textsf{\large 1}}$ $_{\textsf{\large 3}}$};

\node[rectangle with diagonal fill,
diagonal top color=myred,
diagonal bottom color=mygreen,
diagonal from right to left,
draw,font=\sffamily\bfseries,minimum size=1cm, inner sep=0pt, outer
sep=0pt] (s31) [below=0cm of s21] {$^{\textsf{\large 4}}$ $_{\textsf{\large 3}}$};
\node[rectangle with diagonal fill,
diagonal top color=myyellow,
diagonal bottom color=myred,
diagonal from right to left,
draw,font=\sffamily\bfseries,minimum size=1cm, inner sep=0pt, outer
sep=0pt] (s32) [right=0cm of s31] {$^{\textsf{\large 1}}$ $_{\textsf{\large 4}}$};
\node[rectangle with diagonal fill,
diagonal top color=myblue,
diagonal bottom color=myyellow,
diagonal from right to left,
draw,font=\sffamily\bfseries,minimum size=1cm, inner sep=0pt, outer
sep=0pt] (s33) [right=0cm of s32] {$^{\textsf{\large 2}}$ $_{\textsf{\large 1}}$};
\node[rectangle with diagonal fill,
diagonal top color=mygreen,
diagonal bottom color=myblue,
diagonal from right to left,
draw,font=\sffamily\bfseries,minimum size=1cm, inner sep=0pt, outer
sep=0pt] (s34) [right=0cm of s33] {$^{\textsf{\large 3}}$ $_{\textsf{\large 2}}$};

\node[rectangle with diagonal fill,
diagonal top color=mygreen,
diagonal bottom color=myred,
diagonal from right to left,
draw,font=\sffamily\bfseries,minimum size=1cm, inner sep=0pt, outer
sep=0pt] (s41) [below=0cm of s31] {$^{\textsf{\large 3}}$ $_{\textsf{\large 4}}$};
\node[rectangle with diagonal fill,
diagonal top color=myblue,
diagonal bottom color=mygreen,
diagonal from right to left,
draw,font=\sffamily\bfseries,minimum size=1cm, inner sep=0pt, outer
sep=0pt] (s42) [right=0cm of s41] {$^{\textsf{\large 2}}$ $_{\textsf{\large 3}}$};
\node[rectangle with diagonal fill,
diagonal top color=myyellow,
diagonal bottom color=myblue,
diagonal from right to left,
draw,font=\sffamily\bfseries,minimum size=1cm, inner sep=0pt, outer
sep=0pt] (s43) [right=0cm of s42] {$^{\textsf{\large 1}}$ $_{\textsf{\large 2}}$};
\node[rectangle with diagonal fill,
diagonal top color=myred,
diagonal bottom color=myyellow,
diagonal from right to left,
draw,font=\sffamily\bfseries,minimum size=1cm, inner sep=0pt, outer
sep=0pt] (s44) [right=0cm of s43] {$^{\textsf{\large 4}}$ $_{\textsf{\large 1}}$};

\end{tikzpicture}
}
\caption{Superposition}
\end{subfigure}%
\caption{Decomposition in orthogonal Latin squares.}
\label{fig:ex-ols}
\end{figure}

Suppose now that we add a further constraint on the labels: beside representing a permutation of the Cartesian product $[N] \times [N]$, we also require that the two projections are \emph{Latin squares} of order $N$. This means that if we consider only the left (respectively, the right) coordinate of each label, we obtain a square where each number from $1$ to $N$ occurs exactly once in each row and column. This is indeed the case of the square in Figure~\ref{fig:ex-game}, with the Latin squares corresponding to the left and right coordinates depicted in Figure~\ref{fig:ex-ols}.
Pairs of Latin squares of this kind (that is, whose superposition gives a permutation over the Cartesian product of possible entries) are also called \emph{orthogonal}. From the perspective of our game, having a permutation defined by a pair of orthogonal Latin squares implies that no two cells separated by another one can be on the same row or column, over a cycle of length greater than 2.

Although the game described above seems quite detached from any real-world setting at a first glance, there are several applications for it in \emph{cryptography}, particularly in the context of \emph{pseudorandom number generators} (PRNGs). Indeed, the initial cell can be thought of as the seed of a PRNG, with the generated keystream being the sequence of labels encountered along the path where the seed lies. A desirable property in PRNGs is to generate sequences of large periods, which is related to the game's goal of reaching a high score. The fact that the cells' labels define a permutation further ensures that the dynamics of the game is \emph{invertible}, which is useful in the context of \emph{block ciphers} for decryption purposes. Finally, having a permutation defined by a pair of orthogonal Latin squares guarantees a certain amount of \emph{diffusion}, a paramount property for stream and block ciphers to frustrate statistical attacks. As a matter of fact, orthogonal Latin squares correspond to a particular kind of \emph{multipermutation}, which are a useful cryptographic primitive when designing the diffusion layer of a block cipher~\cite{vaudenay94}.

The aim of this paper is to investigate the dynamics of the game above when the two orthogonal Latin squares are defined by \emph{Cellular Automata} (CA). In general, CA represent an attractive approach to design PRNGs for cryptographic purposes, for a twofold reason. First, CA can exhibit a very complex dynamical behavior depending on the underlying local rule, which can be exploited to generate pseudorandom sequences that are hard to predict. Second, the shift-invariance that characterizes CA lends itself to very efficient implementations, both in hardware and software.

Wolfram was the first researcher to propose the use of one-dimensional CA to generate pseudorandom sequences for Vernam-like stream ciphers~\cite{wolfram85}. His idea was to initialize a CA with a random configuration (representing the PRNG's seed) and then iterate the CA for many time steps, taking the trace of the CA's central cell as a pseudorandom keystream. According to Wolfram's claims, the unpredictability of the keystream stemmed from the chaotic dynamics induced by the CA, equipped with rule 30. Unfortunately, later research showed that Wolfram's PRNG is in fact very weak, showing attacks to both recover the initial configuration of the CA~\cite{meier91} and invert its iterations~\cite{koc97}. Martin~\cite{martin08} remarked that some of the weaknesses of this PRNG can be traced back to the poor cryptographic properties of rule 30 when interpreted as a Boolean function. For this reason, more recent works~\cite{formenti14,leporati14} focused on searching larger local rules with a better trade-off of cryptographic properties, using various combinatorial search methods. Still, this research thread does not consider another serious issue when using CA to generate pseudorandom sequences as originally meant by Wolfram: as identified already by Daemen in 1994~\cite{daemen94}, CA always have poor diffusion, due to the local nature of the model that does not allow information to spread very far in a single iteration.

Consequently, studying the dynamics of CA that generate orthogonal Latin squares (also called \emph{orthogonal CA}, or OCA) can be regarded as an alternative approach that starts to address the diffusion issue of classic CA-based PRNGs. In this paper, we perform a preliminary investigation of OCA pairs yielding sequences of maximal period, especially focusing on the case where the underlying local rules are linear.

This work is an extended version of the paper ``\emph{Hip to Be (Latin) Square: Maximal Period Sequences from Orthogonal Cellular Automata}'' presented by the author at CANDAR 2021~\cite{mariot21}. In particular, the new and improved contributions with respect to the conference version are summarized as follows:
\begin{compactenum}
    \item We extend the exhaustive search experiments on the distribution of maximal periods for OCA pairs defined over $\F_2$ up to diameter $6$, leveraging on the combinatorial algorithm described in~\cite{mariot17} to efficiently enumerate the search space. We also extend this investigation to the OCA pairs of diameter $7$ and $8$ constructed in~\cite{mariot17a} by means of evolutionary algorithms. The new results fix an inaccurate claim in the findings of our conference paper, i.e. that the highest maximal period of $2^{2n}-1$ is achievable only by linear OCA pairs. Indeed, the correct results show that there are also maximal period OCA defined by nonlinear rules already from diameter $d=5$.
    \item Leveraging on the theory of \emph{Linear Modular Systems} (LMS), we describe a method to compactly represent the cycle structure of linear OCA pairs. Such a method is based on the computation of the \emph{rational canonical form} of a Sylvester matrix, and allows us to find a simple condition to check whether an OCA pair can attain maximal period. This boils down to verify if the minimal polynomial of the Sylvester matrix is primitive, and it is equivalent (but more efficient, as shown below) to the previous theoretical result of~\cite{mariot21}, where a method to determine the upper bound on the maximal period was given in terms of Lagrange's theorem.
    \item Based on the primitivity check above, we devise a much more time-efficient algorithm to enumerate all linear OCA pairs of maximal period over $\F_2$, implementing it in {\sc Magma}. In this way, we are able to enumerate all such pairs up to diameter $d=16$ in a bit less than one hour, a significant gain over the algorithm used in~\cite{mariot21}, which took almost five days to arrive only up to $d=11$.  The downside of this new algorithm, on the other hand, is its memory usage, with approximately 25GB required to reach $d=16$. Incidentally, we also fix the counts of Table II of~\cite{mariot21}, which were wrong due to an implementation bug, and we provide also the numbers of maximal linear OCA pairs over the alphabet $\F_3$, up to diameter $d=13$.
\end{compactenum}

The rest of this paper is structured as follows. Section~\ref{sec:prelim} covers all preliminary definitions related to CA and orthogonal Latin squares, which are necessary to introduce the main results in the next sections. Section~\ref{sec:dyn-syst} formally defines the dynamical system based on a pair of OCA, and shows the empirical distributions of the maximum periods up to diameter $d=8$. Section~\ref{sec:lin-oca} focuses on linear OCA pairs, providing a characterization of their periods in terms of the rational canonical form of the underlying Sylvester matrix. Next, Section~\ref{sec:enum} presents an improved algorithm to enumerate all linear OCA pairs with maximal period of a given diameter, and reports the results up to $d=16$ and $d=12$ for OCA respectively over the binary and ternary alphabets. Finally, Section~\ref{sec:outro} sums up the key contributions of the paper, and discusses some directions for future research on the subject.

\section{Preliminaries}
\label{sec:prelim}
In this section, we first recall some basic notions about the Cellular Automata (CA) model used in the rest of this paper. We then summarize the main results from the relevant literature related to the construction of orthogonal Latin squares by means of bipermutive CA. As a general notation, for any $n \in \N$ we denote by $[N] = \{1,\cdots, N\}$ the set of all positive integer numbers smaller than or equal to $N$. Further, given $q = p^a$ with $p$ a prime number and $a \in \N$, we use $\F_q$ to denote the finite field of order $q$, with $+$ and $\cdot$ standing respectively for the sum and multiplication operations. In particular, when $q=2$ the sum coincides with the XOR (denoted as $\oplus$), while the multiplication is the logical AND. For any $n \in \N$ we denote the $n$-dimensional vector space over $\F_q$ by $\F_q^n$, with vector sum and multiplication by a scalar induced by the ground field operations in the usual way. Finally, given the finite field $\F_q$, the ring of polynomials in the indeterminate $X$ with coefficients in $\F_q$ is denoted as $\F_q[X]$.

\subsection{Cellular Automata}
\label{subsec:ca}
Cellular automata are one of the oldest natural computing models studied in the literature, and they generally consist of a regular lattice of cells, whose states take values over a finite alphabet. Each cell updates its state in parallel according to the same local rule evaluated over the corresponding neighborhood. Most of the research in this field concerns the \emph{long-term} behavior and properties of CA, which in this case are considered as a particular type of discrete-time dynamical systems. This usually leads to the setting where the cellular lattice is infinite, and a CA can be characterized as a shift-invariant transformation over the \emph{full-shift} space which is uniformly continuous with respect to the Cantor distance~\cite{hedlund69}. In concrete simultations of CA, the lattice must of course be finite, implying that the long-term dynamics is always ultimately periodic. For our paper, we consider a finite model that is even more constrained. In particular, we focus only on the \emph{short-term} behavior of finite CA, often by just considering a single application of the global rule.

Formally, we define the following model of one-dimensional \emph{No-Boundary CA} (NBCA), which is adopted in~\cite{mariot20} to introduce the CA-based construction of orthogonal Latin squares:
\begin{definition}
\label{def:ca}
A \emph{No-Boundary CA} is a vectorial function $F:\F_q^n \to \F_q^{n-d+1}$ defined by a \emph{local rule} $f: \F_q^d \to \F_q$ of \emph{diameter} $d\le n$, where
\begin{equation}
\label{eq:glob-ca}
F(x_1,\cdots,x_n) = (f(x_1,\cdots,x_d),\cdots,f(x_{n-d+1},\cdots,x_n))
\end{equation}
for all $x = (x_1,\cdots,x_n) \in \F_q^n$.
\end{definition}
From a practical perspective, the output coordinate $i \in [n-d+1]$ of a CA is determined by evaluating the local rule $f$ on the \emph{neighborhood} formed by the $i$-th input cell and the $d-1$ cells to its right. The CA is called no-boundary since the local rule is applied only until the coordinate $n-d+1$, as the remaining ones do not have enough neighbors to their right. Clearly, this implies that the global rule of a NBCA can be iterated as long as there are at least $d$ cells remaining in the current cellular array. As we mentioned above, this does not pose an issue since we will be mostly interested in the short-term behavior arising from a single application of the global rule. In this way, we can effectively identify a CA with the vectorial function $F$. For other CA models that also contemplate boundary conditions, we refer the reader to~\cite{kari05}.

CA are usually considered over the binary alphabet, i.e. with $q=2$. In this case, the local rule can be interpreted as a $d$-variable \emph{Boolean function} $f: \F_2^d \to \F_2$, and the most common way to represent it is by means of its \emph{truth table}. In particular, the truth table of $f$ is defined as
\begin{displaymath}
\Omega_f = (f(0,\cdots,0), f(0,\cdots,1), \cdots, f(1,\cdots,1)) \enspace .
\end{displaymath}
Stated otherwise, $\Omega_f$ is the vector that lists the value of $f$ for all $2^d$ input vectors in $\F_2^d$, assuming they are sorted in lexicographic order. The \emph{Wolfram code} of rule $f$ corresponds to the decimal encoding of the truth table $\Omega_f$.

A second common method to uniquely identify a Boolean function is the \emph{Algebraic Normal Form} (ANF). Considering that any element $x$ is \emph{idempotent} over $\F_2$ (i.e., $x^2 = x$), the ANF is the following multivariate polynomial over the quotient ring
$\mathbb{F}_2[x_1,\cdots,x_n]/(x_1^2 \oplus x_1, \cdots, x_n^2 \oplus x_n)$:
\begin{equation}
    P_f(x) = \bigoplus_{I \in 2^{[n]}} a_I \left( \prod_{i \in I} x_i \right) \enspace ,
\end{equation}
where $2^{[n]}$ denotes the power set of $[n] = \{1,\cdots,n\}$. The \emph{algebraic degree} of $f$ is defined in the natural way, i.e. as the number of terms in the largest nonzero monomial of its ANF, or formally as the cardinality of the largest subset $I \in 2^{[n]}$ such that $a_I \ne 0$. Functions of degree at most $1$ are called \emph{affine}, and affine functions whose ANF have a null constant term are called \emph{linear}. When a binary CA is defined by a linear local rule, the next state of each cell is basically an XOR of a subset of cells in its neighborhood.

\begin{figure}[t]
    \centering
    \begin{subfigure}{.5\textwidth}
        \centering
    \begin{tikzpicture}
    [->,auto,node distance=1.5cm, empt node/.style={font=\sffamily,inner
        sep=0pt}, rect
    node/.style={rectangle,draw,thick,font=\bfseries\sffamily,minimum size=0.7cm, inner
        sep=0pt, outer sep=0pt},
        grey node/.style={rectangle,draw,thick,fill=gray!30,font=\bfseries\sffamily,minimum size=0.7cm, inner
        sep=0pt, outer sep=0pt}]
    
    \node [empt node] (c)   {};
    \node [grey node] (c1) [right=0.1cm of c] {1};
    \node [grey node] (c2) [right=0cm of c1] {1};
    \node [grey node] (c3) [right=0cm of c2] {1};
    \node [rect node] (c4) [right=0cm of c3] {0};
    \node [rect node] (c5) [right=0cm of c4] {0};
    \node [grey node] (c6) [right=0cm of c5] {1};
    
    \node [empt node] (f1) [above=0.5cm of c2.east] {{\footnotesize
            $f(0,0,1) = 1$}};
    
    \node [rect node] (p2) [above=0.85cm of c1] {0};
    \node [rect node] (p1) [left=0cm of p2] {0};
    \node [grey node] (p3) [right=0cm of p2] {1};
    \node [grey node] (p4) [right=0cm of p3] {1};
    \node [rect node] (p5) [right=0cm of p4] {0};
    \node [grey node] (p6) [right=0cm of p5] {1};
    \node [rect node] (p7) [right=0cm of p6] {0};
    \node [rect node] (p8) [right=0cm of p7] {0};
    
    \node [empt node] (p7) [below=0.2cm of p1] {};
        \node [empt node] (p7a) [below=0.5cm of c1] {\phantom{M}};
    \node [empt node] (p8) [right=0.07cm of p7] {};
    \node [empt node] (p12) [above=0.5cm of p1.east] {};
    \node [empt node] (p13) [above=0.5cm of p5.east] {};
    \node [empt node] (p14) [above=0.3cm of p13] {\phantom{M}};
    
    \draw [-,thick, mybrace=0.25, decorate, decoration={brace,mirror,amplitude=5pt,raise=0.4cm}]
    (p1.west) -- (p3.east) node [midway,yshift=-0.3cm] {};
    (p1.west) -- (p2.east) node [midway,yshift=0.3cm] {};
    \draw[->,thick,shorten >=0pt,shorten <=0pt,>=stealth] (p8) -- (c1.north);
    \draw[->, draw=white] (p12) edge[bend left] (p13);
      \end{tikzpicture}
    \caption{Local rule evaluation.}
    \label{fig:nbca-a}
    \end{subfigure}%
    \begin{subfigure}{.5\textwidth}
      \begin{tabular}{cc}
            \hline\noalign{\smallskip}
            $x_i,x_{i+1},x_{i+2}$ & $f(x_i,x_{i+1},x_{i+2})$ \\
            \noalign{\smallskip}\hline\noalign{\smallskip}
            000 & 0 \\
            001 & 1 \\
            010 & 0 \\
            011 & 1 \\
            100 & 1 \\
            101 & 0 \\
            110 & 1 \\
            111 & 0 \\
            \hline\noalign{\smallskip}
          \end{tabular}
         \label{fig:nbca-b} 
    \caption{Truth table of rule 90.}
    \label{fig:nbca}
    \end{subfigure}
    \caption{Example of computation in a CA of length $n=8$ equipped with the linear local rule 90 of diameter $d=3$.}
    \label{fig:nbca-ex}
\end{figure}
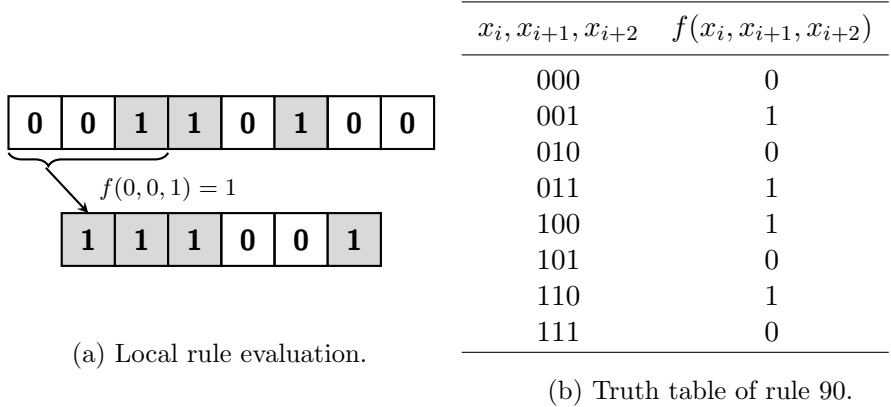

Figure~\ref{fig:nbca-a} depicts an example of CA with $n=8$ input cells, induced by the linear local rule of diameter $d=3$ with ANF $f(x_1,x_2,x_3) = x_1 \oplus x_3$, i.e. only the $i-th$ and $(i+2)-th$ cell in the neighborhood are XORed together. The Wolfram code of this rule is 90, since it corresponds to the decimal encoding of the truth table $(0,1,0,1,1,0,1,0)$, which is reported in Figure~\ref{fig:nbca-a}.

Further information on the ANF of Boolean functions may be found in Carlet's recent book~\cite{carlet21}. In what follows, we will develop our theoretical results for CA over a generic finite field $\F_q$, although our empirical results and examples will mostly refer to the binary case.

\subsection{Orthogonal Latin Squares from Cellular Automata}
\label{subsec:oca}
Let us turn our attention to Latin squares, starting from the following definition:
\begin{definition}
\label{def:ls}
A \emph{Latin square} of order $N \in \N$ is a $N\times N$ matrix $L$ with entries in $[N]$, such that the following two conditions hold:
\begin{compactenum}
\item $L(i,j_1) \neq L(i,j_2)$ for each row coordinate $i \in [N]$ and column coordinates $j_1,j_2 \in [N]$ with $j_1\neq j_1$.
\item $L(i_1,j) \neq L(i_2,j)$ for each column coordinate $j \in [N]$ and row coordinates $i_1,i_2 \in [N]$ with $i_1\neq i_1$.  
\end{compactenum}
\end{definition}
Intuitively, each number from $1$ to $N$ occurs exactly once in each row and in each column of a Latin square of order $N$; equivalently, each row and each column forms a permutation of $[N]$. The concept of orthogonality is defined in terms of the superposition of two Latin squares:
\begin{definition}
\label{def:ols}
Two Latin squares of order $L_1, L_2$ of order $N \in \N$ are called \emph{orthogonal} if for any $(i_1,j_1) \neq (i_2,j_2)$ with $i_1,j_1,i_2,j_2 \in [N]$ it holds that
\begin{displaymath}
(L_1(i_1,j_1), L_2(i_1,j_1)) \neq (L_1(i_2,j_2), L_2(i_2,j_2)) \enspace .
\end{displaymath}
Equivalently, $L_1$ and $L_2$ are orthogonal if the map $H: [N] \times [N] \to [N] \times [N]$ defined as $H(i,j) = (L_1(i,j),L_2(i,j))$ for all $i,j \in [N]$ is bijective.
\end{definition}
From an intuitive point of view, two Latin squares are orthogonal if and only if their \emph{superposition} yields every order pair $(i,j)$ in the Cartesian product $[N] \times [N]$ exactly once. Figure~\ref{fig:ex-ols} in the introductory section of this paper depicts an example of two orthogonal Latin squares of order $4$.

Despite their simple definition, Latin squares spawned a very broad research field, also due to their numerous applications in statistics, cryptography and coding theory. There exist a few known constructions for families of Mutually Orthogonal Latin Squares (MOLS) in the literature, a good account of which can be found in Keedwell and Denes's book~\cite{keedwell15}.

The use of CA to construct orthogonal Latin squares was originally suggested in~\cite{mariot20}, with the original goal of designing a threshold \emph{Secret Sharing Scheme} (SSS). A $(k,n)$-threshold SSS is a protocol that enables a dealer to share a secret value $S$ among a set of $n$ participants, in such a way that at least $k$ participants must combine their respective shares in order to uniquely recover $S$. All coalitions of less than $k$ participants, on the contrary, gain no information on the value of $S$~\cite{shamir79}. It can be shown that families of $n$ MOLS are equivalent to $(2,n)$-threshold SSS (see e.g.~\cite{stinson04}). Most of the SSS based on CA previously published in the literature, on the other hand, feature a \emph{sequential thresold}, meaning that the $k$ shares required to recover the secret must also be adjacent with respect to the order of the participants~\cite{delrey05,mariot14,herranz18}.

The authors of~\cite{mariot20} showed how to generate orthogonal Latin squares with CA, which have later been named \emph{orthogonal CA} (OCA) in~\cite{mariot18}. The construction entails two steps: first, one needs to determine how to define a Latin square from a no-boundary CA. This can be done in a rather natural way by considering CA with \emph{bipermutive local rules}, which we define below:
\begin{definition}
\label{def:bip-rule}
A local rule $f: \F_q^d \to \F_q$ is \emph{bipermutive} if by fixing the leftmost or the rightmost $d-1$ coordinates to any vector $\tilde{x} \in \F_q^{d-1}$, the resulting left and right restrictions $f_{l,\tilde{x}}: \F_q \to \F_q$ and $f_{r,\tilde{x}}: \F_q \to \F_q$ respectively defined for all $x' \in \F_q$ as:
\begin{align}
    \nonumber
    f_{l,\tilde{x}}(x) &= f(\tilde{x}_1, \cdots, \tilde{x}_{d-1}, x')  \\
    \nonumber
    f_{r,\tilde{x}}(x) &= f(x', \tilde{x}_1, \cdots, \tilde{x}_{d-1}) 
\end{align}
are permutations of $\F_q$.
\end{definition}
Remark that for $q=2$ a local rule $f: \F_2^d \to \F_2$ is bipermutive if and only if there exists a $(d-2)$-variable function $g:\F_2^{d-2} \to \F_2$ such that
\[
f(x) = x_1 \oplus g(x_2,\cdots,x_{d-1}) \oplus x_d
\]
for all input vectors $x = (x_1,x_2,\cdots, x_{d-1},x_d) \in \F_2^d$. In other words, rule $f$ depends in a linear way from the leftmost and rightmost variables, since they are independently XORed with a function of the central $d-2$ coordinates. For this reason, CA equipped with bipermutive local rules are also called \emph{quasilinear} in the related literature~\cite{moore97}. Rule 90 used in the example of Figure~\ref{fig:nbca-ex} is bipermutive with $g: \F_2 \to \F_2$ being the zero function.

Consider now a NBCA $F: \F_q^{2(d-1)} \to \F_q^{d-1}$ with a local rule of diameter $d$. Since the input vector is twice the size of the output, we can use the CA to build a square matrix $S_F$ of size $N \times N$ with $N = q^{d-1}$ as follows. Given $x,y \in \F_q^{d-1}$, their concatenation $x\|y \in \F_q^{2(d-1)}$ is used as an input vector for the CA. Then, the output $F(x\|y)$ computed by the CA is the entry of the square $S_F$ where $x$ and $y$ represent respectively the row and column coordinates. From a formal point of view, assume that $\phi: \F_q^{d-1} \to [N]$ is a one-to-one mapping from the vectors of $\F_q^{d-1}$ to the set of the first $N$ positive natural numbers, with $\psi: [N] \to \F_q^{d-1}$ denoting the inverse mapping. Then, for all $i,j \in [N]$ the entry of $S_F$ at row $i$ and column $j$ is defined as:
\begin{equation}
\label{eq:sq-ca}
S_F(i,j) = \phi(F(\psi(i) \| \psi(j))) \enspace .
\end{equation}

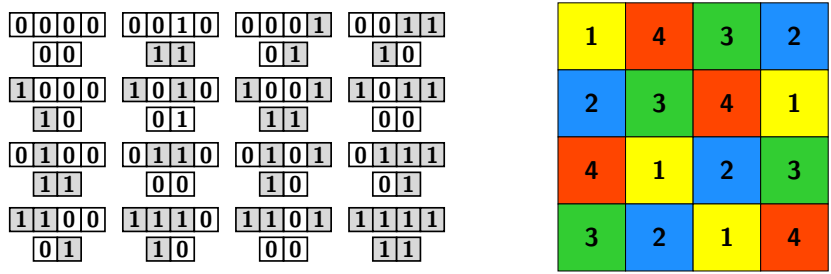
\begin{figure}[ht]
\centering
\begin{subfigure}{.5\textwidth}
\centering
\resizebox{\textwidth}{!}{
\begin{tikzpicture}
[->,auto,node distance=1.5cm,
       empt node/.style={font=\sffamily,inner sep=0pt,minimum size=0pt},
       rect node/.style={rectangle,draw,thick,font=\bfseries\sffamily,minimum size=0.4cm, inner sep=0pt, outer sep=0pt},
       grey node/.style={rectangle,draw,thick,fill=gray!30,font=\bfseries\sffamily,minimum size=0.4cm, inner
        sep=0pt, outer sep=0pt}]
	
        \node [empt node] (e1) {};
	\node [rect node] (i111) [right=0.5cm of e1] {0};
        \node [rect node] (i112) [right=0cm of i111] {0};
        \node [rect node] (i113) [right=0cm of i112] {0};
        \node [rect node] (i114) [right=0cm of i113] {0};
        \node [rect node] (i115) [below=0.1cm of i112] {0};
        \node [rect node] (i116) [right=0cm of i115] {0};

        \node [rect node] (i121) [right=0.3cm of i114] {0};
        \node [rect node] (i122) [right=0cm of i121] {0};
        \node [rect node] (i123) [right=0cm of i122] {1};
        \node [rect node] (i124) [right=0cm of i123] {0};
        \node [grey node] (i125) [below=0.1cm of i122] {1};
        \node [grey node] (i126) [right=0cm of i125] {1};

        \node [rect node] (i131) [right=0.3cm of i124] {0};
        \node [rect node] (i132) [right=0cm of i131] {0};
        \node [rect node] (i133) [right=0cm of i132] {0};
        \node [grey node] (i134) [right=0cm of i133] {1};
        \node [rect node] (i135) [below=0.1cm of i132] {0};
        \node [grey node] (i136) [right=0cm of i135] {1};

        \node [rect node] (i141) [right=0.3cm of i134] {0};
        \node [rect node] (i142) [right=0cm of i141] {0};
        \node [grey node] (i143) [right=0cm of i142] {1};
        \node [grey node] (i144) [right=0cm of i143] {1};
        \node [grey node] (i145) [below=0.1cm of i142] {1};
        \node [rect node] (i146) [right=0cm of i145] {0};

	\node [grey node] (i211) [below=0.7cm of i111] {1};
        \node [rect node] (i212) [right=0cm of i211] {0};
        \node [rect node] (i213) [right=0cm of i212] {0};
        \node [rect node] (i214) [right=0cm of i213] {0};
        \node [grey node] (i215) [below=0.1cm of i212] {1};
        \node [rect node] (i216) [right=0cm of i215] {0};

        \node [grey node] (i221) [right=0.3cm of i214] {1};
        \node [rect node] (i222) [right=0cm of i221] {0};
        \node [grey node] (i223) [right=0cm of i222] {1};
        \node [rect node] (i224) [right=0cm of i223] {0};
        \node [rect node] (i225) [below=0.1cm of i222] {0};
        \node [rect node] (i226) [right=0cm of i225] {1};

        \node [grey node] (i231) [right=0.3cm of i224] {1};
        \node [rect node] (i232) [right=0cm of i231] {0};
        \node [rect node] (i233) [right=0cm of i232] {0};
        \node [grey node] (i234) [right=0cm of i233] {1};
        \node [grey node] (i235) [below=0.1cm of i232] {1};
        \node [grey node] (i236) [right=0cm of i235] {1};

        \node [grey node] (i241) [right=0.3cm of i234] {1};
        \node [rect node] (i242) [right=0cm of i241] {0};
        \node [grey node] (i243) [right=0cm of i242] {1};
        \node [grey node] (i244) [right=0cm of i243] {1};
        \node [rect node] (i245) [below=0.1cm of i242] {0};
        \node [rect node] (i246) [right=0cm of i245] {0};

	\node [rect node] (i311) [below=0.7cm of i211] {0};
        \node [grey node] (i312) [right=0cm of i311] {1};
        \node [rect node] (i313) [right=0cm of i312] {0};
        \node [rect node] (i314) [right=0cm of i313] {0};
        \node [grey node] (i315) [below=0.1cm of i312] {1};
        \node [grey node] (i316) [right=0cm of i315] {1};

        \node [rect node] (i321) [right=0.3cm of i314] {0};
        \node [grey node] (i322) [right=0cm of i321] {1};
        \node [grey node] (i323) [right=0cm of i322] {1};
        \node [rect node] (i324) [right=0cm of i323] {0};
        \node [rect node] (i325) [below=0.1cm of i322] {0};
        \node [rect node] (i326) [right=0cm of i325] {0};

        \node [rect node] (i331) [right=0.3cm of i324] {0};
        \node [grey node] (i332) [right=0cm of i331] {1};
        \node [rect node] (i333) [right=0cm of i332] {0};
        \node [grey node] (i334) [right=0cm of i333] {1};
        \node [grey node] (i335) [below=0.1cm of i332] {1};
        \node [rect node] (i336) [right=0cm of i335] {0};

        \node [rect node] (i341) [right=0.3cm of i334] {0};
        \node [grey node] (i342) [right=0cm of i341] {1};
        \node [grey node] (i343) [right=0cm of i342] {1};
        \node [grey node] (i344) [right=0cm of i343] {1};
        \node [rect node] (i345) [below=0.1cm of i342] {0};
        \node [grey node] (i346) [right=0cm of i345] {1};

	\node [grey node] (i411) [below=0.7cm of i311] {1};
        \node [grey node] (i412) [right=0cm of i411] {1};
        \node [rect node] (i413) [right=0cm of i412] {0};
        \node [rect node] (i414) [right=0cm of i413] {0};
        \node [rect node] (i415) [below=0.1cm of i412] {0};
        \node [grey node] (i416) [right=0cm of i415] {1};

        \node [grey node] (i421) [right=0.3cm of i414] {1};
        \node [grey node] (i422) [right=0cm of i421] {1};
        \node [grey node] (i423) [right=0cm of i422] {1};
        \node [rect node] (i424) [right=0cm of i423] {0};
        \node [grey node] (i425) [below=0.1cm of i422] {1};
        \node [rect node] (i426) [right=0cm of i425] {0};

        \node [grey node] (i431) [right=0.3cm of i424] {1};
        \node [grey node] (i432) [right=0cm of i431] {1};
        \node [rect node] (i433) [right=0cm of i432] {0};
        \node [grey node] (i434) [right=0cm of i433] {1};
        \node [rect node] (i435) [below=0.1cm of i432] {0};
        \node [rect node] (i436) [right=0cm of i435] {0};

        \node [grey node] (i441) [right=0.3cm of i434] {1};
        \node [grey node] (i442) [right=0cm of i441] {1};
        \node [grey node] (i443) [right=0cm of i442] {1};
        \node [grey node] (i444) [right=0cm of i443] {1};
        \node [grey node] (i445) [below=0.1cm of i442] {1};
        \node [grey node] (i446) [right=0cm of i445] {1};
	
\end{tikzpicture}
}
\end{subfigure}%
\begin{subfigure}{.5\textwidth}
\centering
\resizebox{.6\textwidth}{!}{
\begin{tikzpicture}
[->,auto,node distance=1.5cm,
empt node/.style={font=\sffamily,inner sep=0pt,minimum size=0pt},
rect1 node/.style={rectangle,draw,fill=myyellow,font=\sffamily\bfseries,minimum size=1cm, inner
	sep=0pt, outer sep=0pt},
rect2 node/.style={rectangle,draw,fill=myblue,font=\sffamily\bfseries,minimum size=1cm, inner
	sep=0pt, outer sep=0pt},
rect3 node/.style={rectangle,draw,fill=mygreen,font=\sffamily\bfseries,minimum size=1cm, inner
	sep=0pt, outer sep=0pt},
rect4 node/.style={rectangle,draw,fill=myred,font=\sffamily\bfseries,minimum size=1cm, inner
	sep=0pt, outer sep=0pt}]

\node [rect1 node] (s11) {1};
\node [rect4 node] (s12) [right=0cm of s11] {4};
\node [rect3 node] (s13) [right=0cm of s12] {3};
\node [rect2 node] (s14) [right=0cm of s13] {2};

\node [rect2 node] (s21) [below=0cm of s11] {2};
\node [rect3 node] (s22) [right=0cm of s21] {3};
\node [rect4 node] (s23) [right=0cm of s22] {4};
\node [rect1 node] (s24) [right=0cm of s23] {1};

\node [rect4 node] (s31) [below=0cm of s21] {4};
\node [rect1 node] (s32) [right=0cm of s31] {1};
\node [rect2 node] (s33) [right=0cm of s32] {2};
\node [rect3 node] (s34) [right=0cm of s33] {3};

\node [rect3 node] (s41) [below=0cm of s31] {3};
\node [rect2 node] (s42) [right=0cm of s41] {2};
\node [rect1 node] (s43) [right=0cm of s42] {1};
\node [rect4 node] (s44) [right=0cm of s43] {4};

\end{tikzpicture}
}
\end{subfigure}%
\caption{Example of Latin square generated by the CA with local rule $150$.}
\label{fig:r150-sq}
\end{figure}

Eloranta~\cite{eloranta93} and Mariot et al.~\cite{mariot16} independently proved the following sufficient condition for the square $S_F$ to be Latin:
\begin{lemma}
\label{lm:bip-ls}
Let $F: \F_q^{2(d-1)} \to \F_q^{d-1}$ be a NBCA defined by a bipermutive local rule $f: \F_q^d \to \F_q$ of diameter $d$. Then, the square $S_F$ in Equation~\eqref{eq:sq-ca} is a Latin square of order $N = q^{d-1}$.
\end{lemma}
As an example, Figure~\ref{fig:r150-sq} shows the Latin square of order $N=4$ associated to the CA $F: \F_2^4 \rightarrow \F_2^2$ with bipermutive local rule $150$, whose ANF is defined as $f_{150}(x_1, x_2, x_3) = x_1 \oplus x_2 \oplus x_3$. In this case, the mapping $\phi: \F_2^2 \to [N]$ is given by $\phi(0,0) = 1$, $\phi(1,0) \mapsto 2$, $\phi(0,1) = 3$, and $\phi(1,1) = 4$.

After figuring out how Latin squares can be constructed through CA, the next step in the construction is to determine when their superposition yields an orthogonal pair. To this end, the characterization of OCA proved by the authors of~\cite{mariot20} considers bipermutive local rules that are also linear. We have already introduced above linear rules for the binary alphabet, as a particular case of the ANF of Boolean functions. For a generic finite field $\F_q$ a linear rule $f: \F_q^d \to \F_q$ is defined similarly, i.e. as the following linear combination:
\begin{equation}
\label{eq:lin-rule}
f(x_1,\cdots,x_d) = a_1x_1 + \cdots + a_dx_d
\end{equation}
for all $x \in \F_q^d$. Notice that $f$ is bipermutive if and only if the leftmost and rightmost coefficients are not null, that is $a_1 \neq 0$ and $a_d \neq 0$. It is possible to associate a polynomial of degree $n=d-1$ with coefficients in $\F_q$ to a linear rule as follows: 
\begin{equation}
\label{eq:pol-rule}
f \mapsto P_f(X) = a_1 + a_2X + \cdots + a_dX^n
\end{equation}
Hence, we simply use the coefficients of the linear rule reported in Equation~\eqref{eq:lin-rule} as the coefficients of the indeterminate's increasing powers.

The characterization of linear OCA proved in~\cite{mariot20} can be stated as follows:
\begin{theorem}
\label{thm:lin-oca}
Let $F,G: \F_q^{2n} \to \F_q^{n}$ be two NBCA defined by linear bipermutive local rules $f,g: \F_q^d \to \F_q$ of diameter $d$, with $n=d-1$. Then, the two Latin squares of order $N = q^{n}$ generated by F and G are orthogonal if and only if the polynomials $P_f(X), P_g(X) \in \F_q[X]$ of degree $n$ respectively associated to $f$ and $g$ are coprime.
\end{theorem}

Therefore, given two linear bipermutive rules $f,g$ of diameter $d$, it suffices to compute the greatest common divisor of the two associated polynomials $P_f$ and $P_g$. By Theorem~\ref{thm:lin-oca}, the Latin squares $S_f$ and $S_g$ are orthogonal if and only if the GCD of $P_f$ and $P_g$ is $1$.

As an example, the two polynomials over $\F_2$ associated to the local rules $90$ and $150$ of diameter $d=3$ are respectively $P_f(X) = X^2 + 1$ and $P_g(X) = X^2 + X + 1$. Clearly one has $\gcd(P_f,P_g) = 1$ since $P_g$ is irreducible, and thus the corresponding Latin squares $S_F$ and $S_G$ of order $4$ are orthogonal. Indeed, these squares are depicted in the example of Figure~\ref{fig:ex-ols} featured in the Introduction.

\section{Dynamical Systems Induced by OCA}
\label{sec:dyn-syst}
In this section, we start by formally defining the dynamical system to generate pseudorandom sequences by employing a pair of OCA. Next, we present the results of our empirical search experiments on the maximum periods attainable by such a system. These include both an exhaustive search approach up to diameter $d=6$, and an analysis of OCA constructed with Evolutionary Algorithms (EA) for diameters $d=7$ and $d=8$.

\subsection{Formalization and Problem Statement}
\label{subsec:form}
We discussed earlier in Section~\ref{subsec:ca} that a no-boundary CA can be iterated only for a finite number of steps, due to the fact that the size of the cellular array shrinks by $d-1$ cells after each evaluation of the global rule $F$. Hence, although a NBCA equipped with a bipermutive local rule generates a Latin square on account of Lemma~\ref{lm:bip-ls}, it is not possible to use it directly for the generation of pseudorandom sequences. Indeed, a pseudorandom number generator can be viewed as a discrete-time dynamical system $\mathcal{S} = \langle A, f \rangle$ where $A$ is a (finite) set representing the \emph{phase space} of the system, while $f: A \to A$ is an endofunction which maps the current state $s(t) \in A$ at time step $t \in \N$ into the next one $s(t+1) \in A$ at time step $t+1$\footnote{Usually, the general definition of a dynamical system also requires that $A$ is a metric space and that $f$ is continuous with respect to the topology induced by the distance over $A$~\cite{kurka03}. However, since we deal only with the case where the phase space is finite, every update function is trivially continuous.}.

For this reason, the main idea of our pseudorandom generator is to take a \emph{pair} of local rules $f,g: \F_q^d~\to~\F_q$, instead of a single one. Both rules are applied to the same initial configuration $s$ of length $2n = 2(d-1)$, as in the case of orthogonal Latin squares. Hence, one obtains two output vectors $z = F(s)$, $w = G(s)$ of length $n$, generated by the NBCA $F$ and $G$ respectively defined by $f$ and $g$. Next, we construct a new configuration of length $2n$ by concatenating $z$ and $w$. Therefore, the outputs of the NBCA $F, G$ are used respectively as a new row and a new column coordinate, which will in turn point to a new pair of entries given by $F$ and $G$. Considering the superposed representation of the Latin squares generated by $F$ and $G$, this operation can be conceived as starting from the pair of entries occurring at the coordinates indexed by the initial configuration $s$, and using them as the new coordinates where to ``jump'' next (see Figure~\ref{fig:ex-game} in the Introduction).

\begin{figure}[t]
    \centering
    \includegraphics[height=9cm]{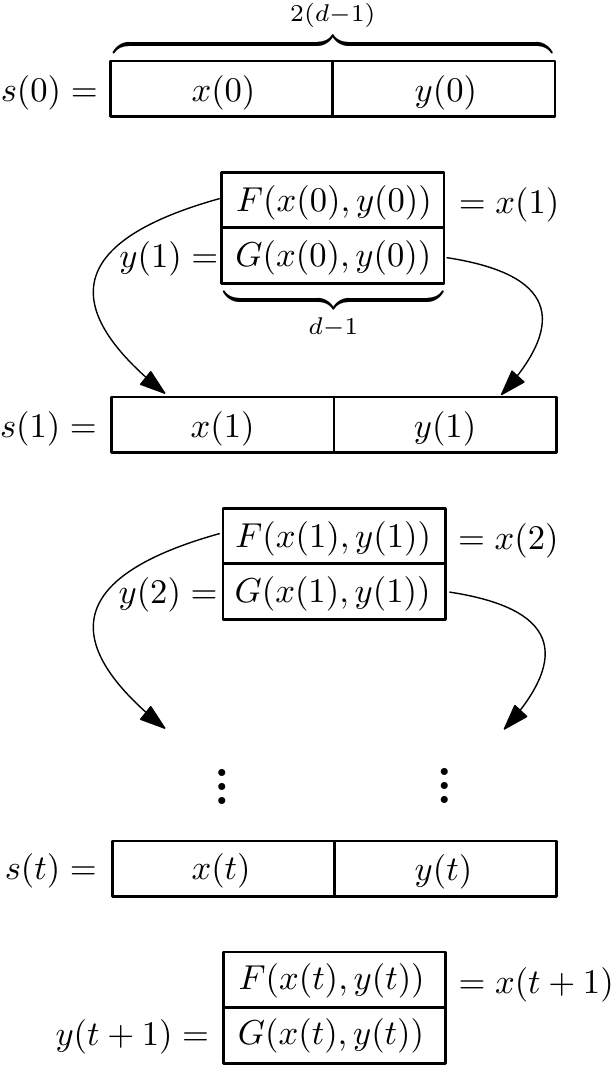}
    \caption{Block diagram for the dynamical evolution of the system starting from the initial state $s(0) = (x(0)\|y(0)) \in \F_q^{2n}$.}
    \label{fig:dynsys}
\end{figure}

We now formally define the dynamical system $\mathcal{S}$ intuitively described above.
\begin{definition}
\label{def:dyn-syst}
Let $d \in \N$ with $d>1$ and $n=d-1$, and let $q=p^a$ be a power of a prime number. 
Additionally, let $F,G: \F_q^{2n} \to \F_q^{n}$ be two OCA defined by bipermutive local rules of diameter $d$. Then, the dynamical system induced by $F$ and $G$ is defined as $\mathcal{S} = \langle A, H \rangle$, where:
\begin{itemize}
\item $A = \F_q^{2n}$, i.e. the phase space is the $2n$-dimensional vector space over $\F_q$.
\item $H: \F_q^{2n} \to \F_q^{2n}$ is the update function defined for all $x, y \in \F_q^{n}$ as:
\begin{equation}
\label{eq:dyn-syst}
H(x\|y) = F(x\|y)\|G(x\|y) \enspace .
\end{equation}
\end{itemize}
\end{definition}
In other words, the state of the system is always separated in two equal-sized parts. When updating the state through $H$, the left part comes from the application of the first NBCA on the whole state in the previous step, whereas the right part is defined analogously as the result of the second NBCA evaluated on the previous state. Figure~\ref{fig:dynsys} depicts the block diagram for the dynamical evolution of the system starting from an initial state $s(0) = x(0)\|y(0)$.
In principle, one could sample the orbit arising from the iteration of Equation~\eqref{eq:dyn-syst} as a pseudorandom sequence, starting from a random initial configuration $s(0) \in \F_q^{2n}$. However, pseudorandom sequences adopted in domains such as cryptography need to satisfy several stringent properties, which implies that randomly selecting the local rule is not a good option. The motivation by which we require that the Latin squares generated by the NBCA $F$ and $G$ are also orthogonal in Definition~\ref{def:dyn-syst} is twofold.

First, as recalled in Section~\ref{subsec:oca}, a pair of orthogonal Latin squares of order $N$ defines a \emph{permutation} over the Cartesian product $[N]\times [N]$. It follows that the update function defined in Equation~\eqref{eq:dyn-syst} is bijective. Thus, the resulting system is \emph{reversible}, or equivalently its trajectories are all disjoint cycles, without transient parts. In practice, reversibility implies that the system can also be run backward in time, by applying the inverse permutation. Such a property is important in certain cryptographic primitives (e.g., SPN block ciphers) where, beside generating pseudorandom sequences, there is also the need of inverting the global state of the cipher to ensure decryption. In the particular setting of OCA, one could invert the system by using the algorithm based on coupled de Bruijn graphs described in~\cite{mariot18}.

Second, orthogonal Latin squares coincide with a particular kind of \emph{Maximum Distance Separable (MDS) codes}, which are of great importance in the design of \emph{diffusion layers} for block ciphers. The reason is that layers based on MDS codes spread the statistical structure of the plaintext over the ciphertext in an optimal way, providing resistance against differential cryptanalysis. In particular, as shown by Vaudenay~\cite{vaudenay94}, the function $H$ defined in Equation~\eqref{eq:dyn-syst} corresponds to a $(2,2)-$\emph{multipermutation}, i.e. any distinct pair of input/output tuples $(x,y, F(x,y), G(x,y))$ and $(x',y',F(x',y'),G(x',y'))$ \emph{cannot agree} on any $2$ coordinates. Thus, such tuples must be at Hamming distance at least $3$.

The aim of this work is to investigate the cyclic structure of the dynamical system $\mathcal{S}$, paying particular attention to cycles of maximal period. Given a state $s \in \F_q^{2n}$, the (minimum) \emph{period} of $s$ under $\mathcal{S}$ is the smallest positive integer $p$ such that $H^{p}(s) = s$. In other words, $p$ is the smallest number of iterations of $H$ after which the state of the system returns to the initial condition $s$. Pseudorandom sequences with very large periods are usually sought in cryptography especially in the context of stream ciphers~\cite{stinson18}. Indeed, if a pseudorandom sequence used as a keystream has a shorter period than the plaintext length, an adversary can mount certain attacks based on frequency analysis. Ideally, the dynamics of a pseudorandom generator used in cryptography should be composed of a single large cycle that visits (in a non-trivial and unpredictable way) all states in the phase space.

We conclude this section by formally stating the problem addressed in the rest of this paper:
\begin{problem}
\label{pb:stat}
Let $d \in \N$ and $q$ be a power of a prime number, and let $n=d-1$. What is the largest period attainable by the system $\mathcal{S} = \langle \F_q^{2n}, H \rangle$, with $H$ defined as in Equation~\eqref{eq:dyn-syst}, when $F$ and $G$ are OCA?
\end{problem}

\subsection{Empirical Search Results}
\label{subsec:exhaustive}

We begin our study of the periods of OCA by performing an empirical search, focusing on the case of the binary alphabet, i.e. $q=2$. The number of all Boolean functions of $d$ variables is $2^{2^d}$, since one can assign either 0 or 1 to each of the $2^d$ input vectors in the truth table. This prevents any exhaustive search already for $d>5$ variables. However, concerning Problem~\ref{pb:stat} we are only interested in those dynamical systems $\mathcal{S}$ defined by two NBCA of diameter $d$ whose local rules are bipermutive Boolean functions. A bipermutive function of $d$ variables is effectively defined by the generating function of the $d-2$ central cells, and thus the total number  of bipermutive functions to enumerate is $2^{2^{d-2}}$. Since we consider pairs of bipermutive local rules, we have that the search space is composed of $2^{2^{d-2}} \times 2^{2^{d-2}} = 2^{2^{d-1}}$ feasible solutions. This allows to stretch the exhaustive search approach up to $d=6$, since in that case we have at most $2^{2^{6-1}} \approx 4.3 \cdot 10^9$ pairs to enumerate. Further, Mariot et al.~\cite{mariot17} gave a necessary condition on the local rules of two OCA, showing that their truth tables must be \emph{pairwise balanced}. This allows us to further reduce the search space to about $6.3 \cdot 10^7$ pairs, using the combinatorial algorithm described in~\cite{mariot17}.

Beyond diameter $d=6$ exhaustive search becomes unfeasible. For this reason, to expand the scope of our empirical search (especially with respect to our previous conference paper~\cite{mariot21}), we also considered two samples of OCA pairs of diameters $d=7$ and $d=8$. Such samples are taken from the paper~\cite{mariot17a}, where the authors employed \emph{Genetic Algorithms} (GA) and \emph{Genetic Programming} (GP) to construct OCA pairs, and they are composed respectively of $68$ pairs for $d=7$ and $50$ pairs for $d=8$.

For each pair $f,g: \F_2^d \to \F_2$ of bipermutive local rules considered in our empirical search, we need to perform the following steps:
\begin{compactenum}
    \item Check if the Latin squares generated by the NBCA $F,G: \F_2^{2n} \to \F_2^n$ respectively defined by $f$ and $g$ are orthogonal. Of course, this step is optional for the samples of $d=7$ and $d=8$, since there we already know that all pairs induce OCA.
    \item If the two NBCA are orthogonal, compute the cycle decomposition of the dynamical system $\mathcal{S} = \langle \F_2^{2n}, H\rangle$ with $H$ defined as in Equation~\eqref{eq:dyn-syst}.
    \item Find the length of the largest cycle(s) in $\mathcal{S}$.
\end{compactenum}
It makes sense to start our empirical search from diameter $3$: in fact, there are no OCA pairs of diameter $2$, since there do not exist orthogonal Latin squares of order $2^{2-1} = 2$ in general, be them induced by CA or not. For diameter $d=3$, a total of $8$ OCA pairs result from the search over all 16 pairs of bipermutive rules. All these OCA pairs yielded the same cycle decomposition structure, i.e. one fixed point and a single maximal cycle of length $15$. This is expected, since for $d=3$ only linear OCA pairs exist, and they are all equivalent by three symmetry relations observed in~\cite{mariot17}, namely \emph{swap}, \emph{complement} and \emph{reflection}. In particular, the swap symmetry changes the order of the local rules in a pair, the complement negates the truth tables of both rules, and reflection evaluates them on the input in reversed order. Each of these symmetries is an equivalence relation which halves the search space of local rules pairs. Hence, the number all OCA pairs can actually be divided by $8$, meaning that there exists only a single OCA pairs of diameter $d=3$ up to swap, complement and reflection. This explains why the 8 pairs mentioned above all exhibit the same dynamics.

\begin{figure}[t]
    \centering
    \includegraphics[width=\textwidth]{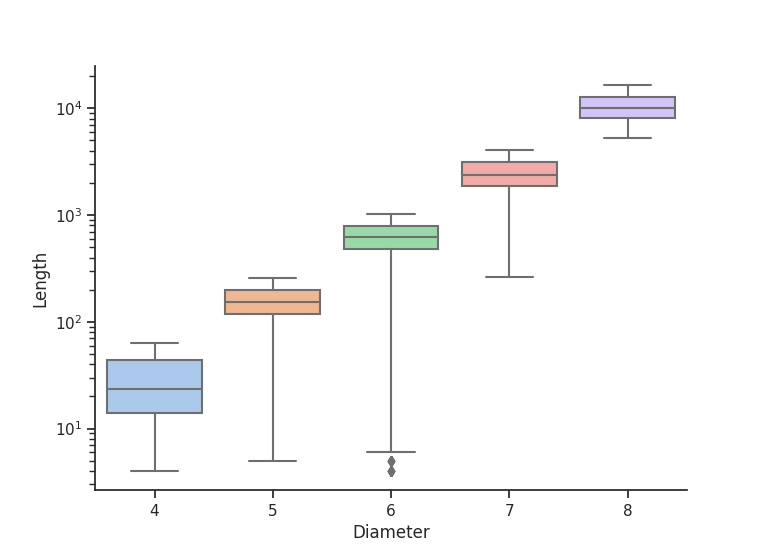}
    \caption{Distribution of maximum cycle lengths for OCA of diameters $4 \le d \le 8$.}
    \label{fig:dist-d4-d8}
\end{figure}

\begin{figure}[!h]
\begin{subfigure}{\textwidth}
    \centering
    \includegraphics[width=0.9\textwidth]{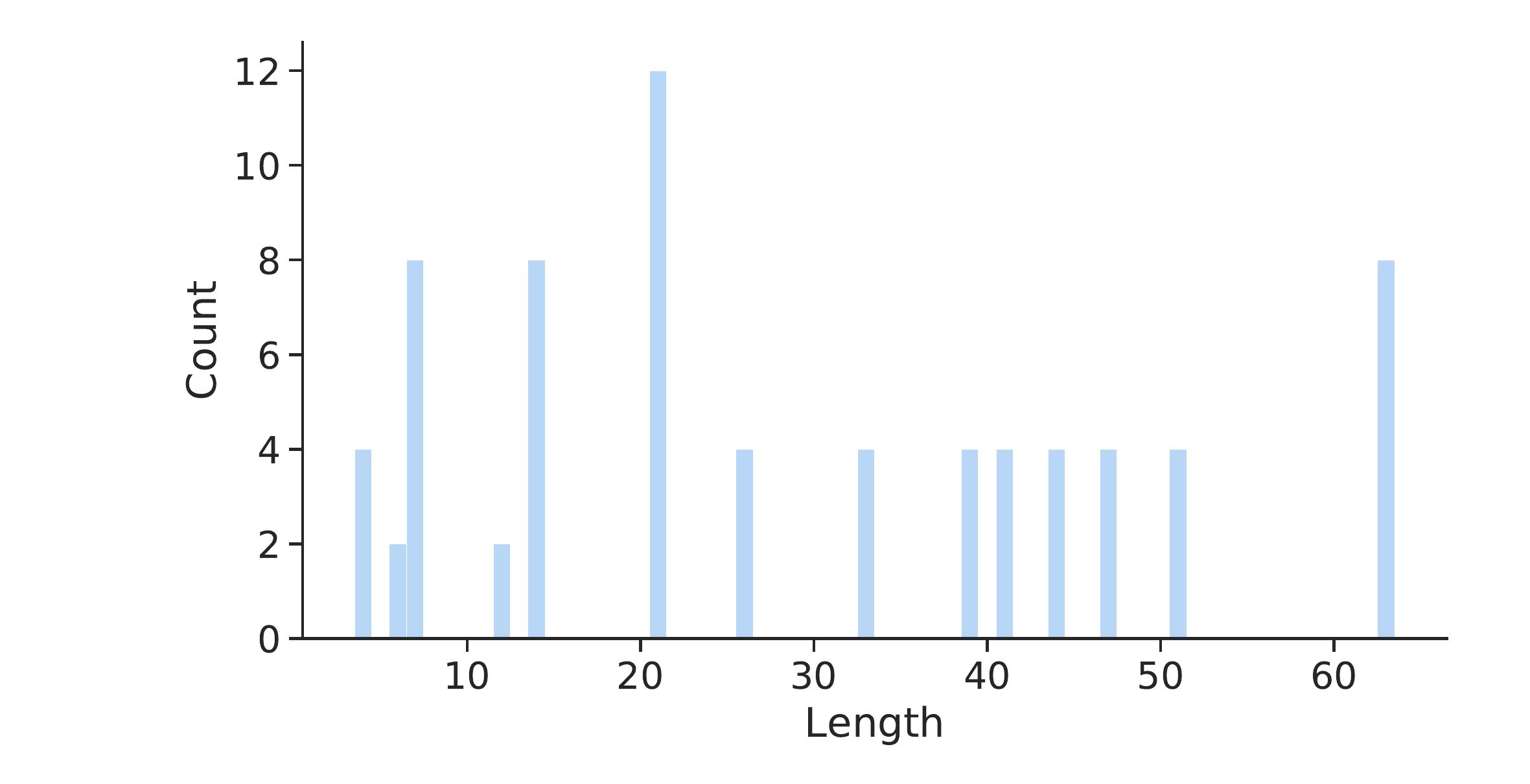}
    \caption{$d=4$}
    \label{fig:dist-d4}
\end{subfigure}
    
\begin{subfigure}{\textwidth}
    \centering
    \includegraphics[width=\textwidth]{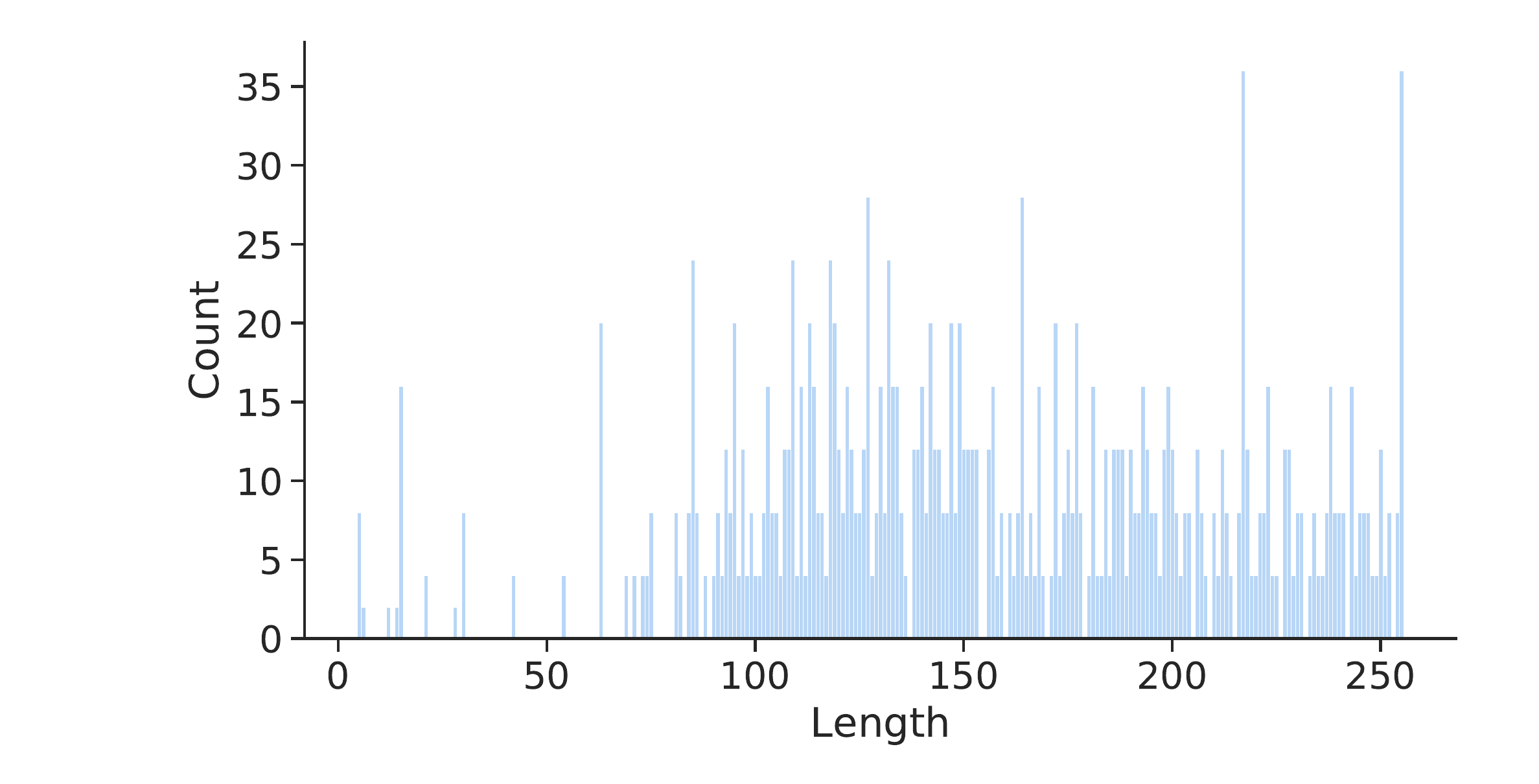}
    \caption{$d=5$}
    \label{fig:dist-d5}
\end{subfigure}
\caption{Distribution of maximum cycle lengths for OCA of diameter $d=4,5$.}
\label{fig:dist-d4-d5}
\end{figure}

We refer to the boxplots in Figure~\ref{fig:dist-d4-d8} for a general outlook of the distributions of maximal cycle lengths for diameters $4\le d \le 8$.
Figures~\ref{fig:dist-d4} and~\ref{fig:dist-d5} depict more in detail the distributions of diameters $d=4$ and $d=5$ as histograms. We omitted the histograms for the remaining diameters since they could not be displayed properly, due to either too dense (for $d=6$) or too sparse (for $d=7$ and $d=8$) distributions.
As a general remark, one can notice that the distributions up to $d=6$ all have a very small minimum value. This is reasonable, since in all those cases we were able to perform an exhaustive search, meaning that we are considering the \emph{complete} distributions, instead of a sample. Hence, our exhaustive search finds several OCA pairs characterized by many cycles of small length, or even by many fixed points. In any case, it is interesting to observe that the interquartile range is always compressed towards the maximum value, meaning that the great majority of OCA pairs have a large maximum period. This trend is confirmed also for the distributions of $d=7$ and $d=8$. Indeed, here the minimum values are way above those of $d\le 6$, which suggests that the GA and GP proposed in~\cite{mariot17a} are able to sample OCA pairs with large maximum cycle lengths. A third interesting remark, moreover, is that \emph{the largest maximum cycle length observed in our experiments is $2^{2n}-1$}. In other words, we found no OCA pairs giving a ``pure cycle'' of length $2^{2n}$ which generates the whole phase space $\F_2^{2d}$. The best possible setting seems always to be the case where an OCA visit all cells in the superposed squares except one, that represents a fixed point.

For each diameter $2 \le d \le 6$, Table~\ref{tab:sizes} reports the results of our exhaustive search concerning those OCA pairs reaching a maximum cycle length of $2^{2n}-1$. In particular, the first six columns from left to right report respectively the diameter $d$ and $n=d-1$, the order of the corresponding Latin squares $2^n$, the number of bipermutive local rules $\mathcal{B}_d = 2^{2^{d-2}}$, the number of ordered pairs  $\mathcal{B}_d^2$ visited by our exhaustive search algorithm, and the number of pairs which generate OCA $\mathcal{OCA}_d$ (taken from~\cite{mariot17}). Finally, the last three columns report the total number $\#m\mathcal{OCA}_d$ of OCA pairs having a maximum cycle length of $2^{2n}-1$, and then their classification in nonlinear and linear pairs, respectively denoted as $\#m\mathcal{NOCA}_d$ and $\#m\mathcal{LOCA}_d$.
\begin{table}[t]
\centering
\caption{Exhaustive search results for OCA pairs of diameter $2 \le d \le 6$.}
\begin{tabular}{lllllllll}
\hline\noalign{\smallskip}
$d$ & $n$ & $2^{n}$ & $\mathcal{B}_d$ & $\mathcal{B}^2_d$ & $\mathcal{OCA}_d$ & $\#m\mathcal{OCA}_d$ & $\#m\mathcal{NOCA}_d$ & $\#m\mathcal{LOCA}_d$ \\
\noalign{\smallskip}\hline
2 & 1 & 2 & 2 & 4 & 0 & 0 & 0 & 0 \\
3 & 2 & 4 & 4 & 16 & 8 & 8 & 0 & 8 \\
4 & 3 & 8 & 16 & 256 & 72 & 8 & 0 & 8 \\
5 & 4 & 16 & 256 & 65536 & 1704 & 36 & 12 & 24 \\
6 & 5 & 32 & 65536 & $6.3 \cdot 10^7$ & 533480 & 1968 & 1840 & 128 \\
\hline
\end{tabular}
\label{tab:sizes}
\end{table}
Remark that the numbers in the last four columns of Table~\ref{tab:sizes} are not normalized up to the three symmetry relations mentioned above. The values given in the seventh column of the table show that the number of OCA pairs with a maximum cycle length of $2^{2n}-1$ represent a very small fraction of all OCA pairs, which moreover becomes even smaller as the diameter increases. Further, contrary to what we reported in our previous conference work~\cite{mariot21}, \emph{there do exist nonlinear OCA pairs of maximum cycle length $2^{2n}-1$}. Our previous claim was wrong (i.e., we reported that only linear pairs achieved the maximum observed cycle length) due to a bug in the Java code used to decompose the cycles of OCA pairs, which has been located and fixed after our conference paper was published in CANDAR 2021. Indeed, by extending the search to $d=6$ (while in~\cite{mariot21} we arrived at $d=5$), one can even see that the proportion of linear OCA is quite small compared to that of nonlinear OCA. Nevertheless, in what follows we focus on the linear OCA  since in that case it is possible to use results from linear algebra over finite fields to give a precise characterization of their cycle structures.

\section{Periods of Linear OCA}
\label{sec:lin-oca}
We now focus on the linear case, describing a method to completely determine the cycle structure of a pair of linear OCA. This improves on the previous results of our conference paper~\cite{mariot21}, where only an upper bound on the maximal period of linear OCA was given. As it often happens when studying the behavior of dynamical systems governed by a linear transformation, our method leverages on linear algebra methods, and in particular on the theory of \emph{Linear Modular Systems} (LMS). A good overview of the results that we use in this section can be found in Lidl and Niederreiter's book on finite fields~\cite{lidl97}.

Let $f,g: \F_q^d \to \F_q$ be two linear bipermutive local rules of diameter $d$. Following the notation recalled in Section~\ref{sec:prelim}, we assume that the linear combinations defining $f$ and $g$ are respectively given by the two vectors $a = (a_1, \cdots, a_d) \in \F_q^d$ and $b = (b_1, \cdots, b_d) \in \F_q^d$, where $a_1,b_1,a_d,b_d$ are all nonzero to ensure bipermutivity. In particular, we assume that $a_d=b_d=1$ to obtain monic polynomials, which simplifies our calculations. Therefore, suppose that $P_f(X), P_g(X) \in \F_q[X]$ are the monic polynomials of degree $n=d-1$ and nonzero constant term associated to $f$ and $g$. By Theorem~\ref{thm:lin-oca} $f$ and $g$ induce a pair of OCA if and only if their polynomials $P_f(X)$ and $P_g(X)$ are relatively prime. As proved in~\cite{mariot20}, this characterization stands on the fact that the transformation which associates the CA input configuration $x\|y \in \F_q^{2n}$ to the output $F(x\|y)\|G(x\|y)$ is defined by the following $2n \times 2n$ \emph{Sylvester matrix}:

\begin{equation}
    \label{eq:sylv-matr}
    M_{f,g} =
    \begin{pmatrix}
      a_1    & \cdots & a_{d} & 0 & \cdots & \cdots & \cdots & \cdots & 0 \\
      0      & a_1    & \cdots  & a_{d} & 0 & \cdots & \cdots & \cdots & 0 \\
      \vdots & \vdots & \vdots & \ddots  & \vdots & \vdots & \vdots & \ddots & \vdots \\
      0 & \cdots & \cdots & \cdots & \cdots & 0 & a_1 & \cdots & a_{d} \\
      b_1    & \cdots & b_{d} & 0 & \cdots & \cdots & \cdots & \cdots & 0 \\
      0      & b_1    & \cdots  & b_{d} & 0 & \cdots & \cdots & \cdots & 0 \\
      \vdots & \vdots & \vdots & \ddots  & \vdots & \vdots & \vdots & \ddots & \vdots \\
      0 & \cdots & \cdots & \cdots & \cdots & 0 & b_1 & \cdots & b_{d} \\
\end{pmatrix} \enspace .
\end{equation}
In particular, the two rules generate a pair of OCA if and only if the transformation $M_{f,g}\cdot (x,y)^\top$ is bijective, or equivalently if and only if $M_{f,g}$ is invertible. It is a well known fact that the determinant of a Sylvester matrix---also called the \emph{resultant}---is not null if and only if $P_f(X)$ and $P_g(X)$ do not have any factor in common~\cite{gelfand08}. Hence, the authors' focus in~\cite{mariot20} was to count the number of linear OCA pairs by counting the number of invertible Sylvester matrices defined by linear bipermutive rules, or equivalently the number of pairs of coprime polynomials with degree $n$ and nonzero constant term over $\F_q$.

As it usually happens when dealing with a dynamical system whose updating function is described by a matrix, the $t$-th iterate of the system $\mathcal{S}$ defined in Section~\ref{subsec:form} consists of multiplying the $t$-th power of the Sylvester matrix $M_{f,g}$ by the initial state vector, as shown in the next lemma:
\begin{lemma}
\label{lm:t-iter}
Given $d \in \N$ and $n=d-1$, let $\mathcal{S} = \langle \F_q^{2n}, H \rangle$ be the dynamical system defined by the update function in Equation~\eqref{eq:dyn-syst}, where the CA $F,G: \F_q^{2n} \to \F_q^n$ are defined by two bipermutive linear rules $f,g: \F_q^d \to \F_q$ of diameter $d$, with coprime associated polynomials $P_f(X), P_g(X) \in \F_q[X]$. Then, for any initial state $s(0) = x(0)\|y(0) \in \F_q^{2n}$, the state of $\mathcal{S}$ at time $t \in \N$ is given by:
\begin{equation}
\label{eq:t-iter}
s(t) = x(t)\|y(t) = M_{f,g}^t \cdot s(0) = M_{f,g}^t \cdot (x(0) \| y(0))^\top \enspace .
\end{equation}
\begin{proof}
We proceed by induction on $t \in \N$. The base case $t=1$ corresponds to the observation above about Theorem~\ref{thm:lin-oca}: a single application of the map $H: \F_2^{2n} \to~\F_2^{2n}$ defined in Equation~\eqref{eq:dyn-syst} corresponds to the matrix-vector multiplication $M_{f,g} \cdot (x(0)\|y(0))^\top$. Let us assume now that the claim is valid for any $t \in \N$, and consider the case $t+1$: this is equivalent to iterating $H$ for $t+1$ steps starting from $s(0)$, which can be written equivalently as the composition of $H$ with its $t$-th iterate $H^t$:
\begin{equation}
\label{eq:t-iter2}
s(t+1) = H^{t+1}(s(0)) = H \circ H^t(s(0)) \enspace .
\end{equation}
By induction hypothesis, we know that $H^{t}(s(0)) = M_{f,g}^t \cdot s(0)^\top$, and that a single application of $H$ amounts to multiplying $M_{f,g}$ with the current state vector. Hence, we can rewrite Equation~\eqref{eq:t-iter2} as follows:
\begin{equation}
\label{eq:t-iter3}
H^{t+1}(s(0)) = H \circ H^t(s(0)) = M_{f,g} \cdot (M_{f,g}^t \cdot s(0)^\top)^\top \enspace ,
\end{equation}
from which we conclude that
\[
s(t+1) = M_{f,g}^{t+1} \cdot s(0)^\top = M_{f,g}^{t+1} \cdot (x(0), y(0))^\top .
\]
\end{proof}
\end{lemma}

Concerning Problem~\ref{pb:stat}, Lemma~\ref{lm:t-iter} implies that the maximum length of the cycles in system $\mathcal{S}$ are bounded above by the \emph{order} of the associated Sylvester matrix $M_{f,g}$, considered as an element of the \emph{general linear group} $GL(2n, \F_q)$. The general linear group $GL(2n, \F_q)$ is defined as the set of all invertible matrices of size $2n \times 2n$ with entries in $\F_2$, equipped with matrix multiplication as a group operation. Indeed, the orthogonality requirement forces $M_{f,g}$ to be invertible, and Lemma~\ref{lm:t-iter} establishes that the $t$-th iterate of the transformation $H$ corresponds to the $t$-th power of such matrix. Thus, determining the upper bound for the maximum cycle length is equivalent to finding the minimum $t \in \N$ such that $M_{f,g}^t = I_{2n}$, i.e. the $t$-th power of $M_{f,g}$ which transforms it into the identity matrix of order $2n$. This is, in turn, equivalent to determining the order of the cyclic subgroup generated by $M_{f,g}$ in $GL(2n, \F_q)$. It is a well-known fact (see e.g.~\cite{jacobson85,mullen13}) that the order of the general linear group $GL(2n, \F_q)$, or equivalently its cardinality, is equal to:
\begin{equation}
\label{eq:gl}
\#GL(2n,\F_q) = (q^{2n}-1) (q^{2n} - q) (q^{2n} - q^2) \cdots (q^{2n} - q^{2n-1}) \enspace .
\end{equation}

Let us now recall \emph{Lagrange's theorem}~\cite{gallian12}: \emph{the order of any subgroup $H \le G$ of a finite group $G$ must divide the order of $G$}. This means that the order of the cyclic subgroup generated by the Sylvester matrix can only be a divisor of $\#GL(2n,\F_q)$ as defined in Equation~\eqref{eq:gl}. Moreover, we know that the maximum period reachable by a pair of OCA can be at most $q^{2n}$, due to the fact that the phase space $\F_2^{2n}$ of $\mathcal{S}$ is composed of $q^{2n}$ elements, and the null vector is always a fixed point (because the underlying system is linear). Thus, we have concluded that the order of the Sylvester matrix can be at most $q^{2n}-1$, therefore obtaining an upper bound for the maximum cycle length achievable by a pair of linear OCA. To summarize, we have proved the following result:
\begin{lemma}
\label{thm:max}
Let $d \in \N$, $n=d-1$ and $\mathcal{S} = \langle \F_q^{2n}, H \rangle$ be the dynamical system where $H$ is defined as in Equation~\eqref{eq:dyn-syst}, with OCA $F,G: \F_q^{2n} \to \F_q^n$ generated by a pair of linear bipermutive rules $f,g: \F_q^d \to \F_q$. Then, the period $p$ of any state $s \in \F_q^{2n}$ is at most $p \le q^{2n} - 1$.
\end{lemma}

It is important to stress that the bound above is not always reached. Indeed, it might be the case that even though the Sylvester matrix has maximum order $q^{2n}-1$, the cycle structure of two linear OCA is characterized by shorter periods. In particular, assume that the system $\mathcal{S} = \langle \F_q^{2n}, H \rangle$ has cycles of periods $t_1, \cdots, t_k$. Then, the order of the Sylvester matrix $M_{f,g}$ is actually the least common multiple of $t_1, \cdots, t_k$. As a matter of fact, assume that $t$ is the order of $M_{f,g}$: we have that $M_{f,g}^t\cdot s^\top$ for any state $s \in \F_q^{2n}$. Thus, $t$ must be a multiple of $l=\textrm{lcm}(t_1,\cdots,t_k)$. Moreover, $(A^l - I)\cdot s^\top = 0$ for all $s \in \F_q^{2n}$, where $I$ denotes the identity matrix. Therefore, we obtain that $A^l = I$, which means that $l \ge t$, and thus $t = \textrm{lcm}(t_1,\cdots, t_k)$.

To give a more precise characterization of the cycle structure of the system $\mathcal{S}$ in the linear case, we introduce the following \emph{cycle sum} notation following~\cite{lidl97}:

\begin{equation}
\label{eq:cyc-sum}
\sum(\mathcal{S}) = (n_1, t_1) + (n_2, t_2) + \cdots + (n_k, t_k) \enspace .
\end{equation}
This is a formal sum which indicates that $\mathcal{S}$ has $n_i$ cycles of length $t_i$, for all $i \in [k]$. A summand in~\eqref{eq:cyc-sum} is also called a \emph{cycle term}.

Recall that the \emph{characteristic polynomial} of a square matrix $A$ over $\F_q$ is defined as the determinant of $XI - M$, while the \emph{minimal polynomial} of $A$ is the monic polynomial $m(X) \in \F_q[X]$ of smallest degree such that $m(M)$ is the zero matrix. A monic polynomial $g(X) = X^k + a_{k-1}X^{k-1} + \cdots a_1X + a_0 \in \F_q[X]$ is the characteristic and minimal polynomial of its associated \emph{companion matrix} $M(g(X))$. In particular, the characteristic polynomial of any square matrix $A$ over $\F_q$ is the product of its \emph{elementary divisors} $g_1(X),\cdots,g_r(X)$, and the \emph{rational canonical form} of $A$ is the matrix $A^*$ defined as:
\begin{equation}
\label{eq:rcf}
A^* = 
\begin{pmatrix}
M(g_1(X)) & 0 & \cdots & 0 \\
0 & M(g_2(X)) & \cdots & 0 \\
\vdots & \vdots & \ddots & \vdots \\
0 & 0 & \cdots & M(g_r(X))
\end{pmatrix}
\enspace ,
\end{equation}
where $M(g_i(X))$ denotes the companion matrix of the elementary divisor $g_i(X)$. The two matrices $A$ and $A^*$ are related by the equation $A = P^{-1}AP$, where $P$ is an invertible matrix over $\F_q$.

The cycle structure of a linear modular system defined by a transition matrix $A$ can be expressed in terms of the orders of the elementary divisors occurring in its rational canonical form $A^*$. In particular, given a linear OCA pair defined by a nonsingular Sylvester matrix $M_{f,g} \in GL(2n, \F_q)$, the cycle sum of the corresponding dynamical system $\mathcal{S} = \langle \F_q^{2n}, H \rangle$ can be determined using the following procedure described in~\cite{lidl97}:
\begin{enumerate}
\item Determine the elementary divisors $g_1(X),\cdots,g_r(X)$ of the Sylvester matrix $M_{f,g}$, where $g_i(X) = f_i(X)^{m_i}$ with $f_i(X)$ monic and irreducible over $\F_q$ for all $i \in [r]$.
\item Determine the orders $t_1^{(i)} = \textrm{ord}(f_i(X))$ of the polynomials $f_i(X)$.
\item Compute the orders $t_h^{(i)} = \textrm{ord}(f_i(X)^h)$ for $i \in [r]$  and $h \in [m_i]$.
\item Find the cycle sum $\sum(\mathcal{S}_i)$ of the system defined by the elementary block $M(g_i(X))$, for $i \in [r]$, using Theorem 9.96 in~\cite{lidl97}.
\item Determine the cycle sum of the whole system $\mathcal{S}$ as the product of the cycle sums $\sum(\mathcal{S}_i)$, for $i \in [r]$. 
\end{enumerate}
The details to compute the product of cycle sums in the last step of the procedure are omitted for the sake of brevity, but can be found in~\cite{lidl97}.

\section{Enumeration Algorithms and Results}
\label{sec:enum}
Given a pair of linear OCA, the procedure described at the end of the previous section can be used to completely determine the cycle structure of the associated dynamical system $\mathcal{S}$. In this last section, we are interested in determining when the order of the Sylvester matrix $M_{f,g}$ is \emph{exactly} the maximum allowed by Lemma~\ref{thm:max}, i.e. $q^{2n}-1$. Recall that an irreducible polynomial $p(X) \in \F[X]$ of degree $d$ is called \emph{primitive} if it is a generator of the multiplicative group of the extension field $\F_{q^d}$. Then, one has the following result (see e.g.~\cite{ghorpade11} for a proof):
\begin{theorem}
\label{thm:prim}
Let $A \in GL(2n,\F_q)$ be a $2n \times 2n$ nonsingular matrix over $\F_q$, and let $t$ be the order of $A$, i.e. the smallest $t \in \N$ such that $A^t = I$. Then, $t = q^{2n}-1$ if and only if its minimal polynomial $m_A(X)$ is primitive.
\end{theorem}
Hence, to enumerate all linear OCA pairs whose associated Sylvester matrix has maximum order $q^{2n}-1$, we can determine its minimal polynomial and check whether it is primitive. This strategy is summarized in the following procedure:
\begin{itemize}
\item Set $n=d-1$.
\item For each pair of polynomials $P_f(X), P_g(X) \in \F_q[X]$ with degree $n$ and nonzero constant term do:
  \begin{itemize}
    \item if $\gcd(P_f(X), P_g(X)) = 1$ then:
    \begin{itemize}
        \item Determine the minimal polynomial $m(X)$ of $M_{f,g}$
        \item If $m(X)$ is primitive print the pair $P_f(X), P_g(X)$
    \end{itemize}
  \end{itemize}
\end{itemize}
Remark that this enumeration algorithm is different from the one proposed in our previous conference paper~\cite{mariot21}: there, we employed a different method to determine the order of the Sylvester matrix, namely relying on Lagrange's theorem to check only the divisors of the order of $GL(2n,\F_q)$.

We implemented the procedure above in {\sc Magma}, and applied it to enumerate Sylvester matrices of maximum order $q^{2n}-1$ for $q=2$. In particular, this improved enumeration algorithm turned out to be much more efficient than our previous version, since we managed to enumerate all such matrices for linear OCA pairs up to diameter $d=16$ in a bit less than an hour, using a 64-bit Linux machine with a 16-core AMD Ryzen processor running at 3.5 GHz and 48 GB of RAM. In contrast, our previous algorithm based on Lagrange's theorem implemented in Java took almost 5 days to enumerate all such pairs only up to $d=11$, using the same machine. The bottleneck of our new algorithm, on the other hand, becomes the memory: the check of primitivity is likely the step where {\sc Magma} consumes the most memory, and for $d=16$ it reached $25$ GB. We did not manage to go further since for the next instance of $d=17$ we ran out of memory. Beside this experiment, we also applied our improved algorithm to enumerate invertible Sylvester matrices over a ternary alphabet, i.e. with $q=3$. In this case, the time becomes again the bigger bottleneck before the memory does: the enumeration for diameter $d=14$ did not finish within 10 days of computation, hence we stopped at $d=13$.

Table~\ref{tab:enum} reports the numbers obtained from the two experiments described above. The third and fourth column give for each diameter the maximum possible order for Sylvester matrices of size $2n$ respectively over $\F_2$ and $\F_3$. The fifth and the sixth column, likewise, report the number of Sylvester matrices reaching those orders.
\begin{table}[t]
\centering
\caption{Number of invertible $2n \times 2n$ Sylvester matrices of maximum order over $\F_q$, with $q=2,3$.}
\begin{tabular}{cccccc}
\hline\noalign{\smallskip}
$d$ & $n$ & $2^{2n}-1$ & $3^{2n}-1$ & $\mathcal{M}_2$ & $\mathcal{M}_3$ \\
\noalign{\smallskip}\hline
2 & 1 & 3 & 80 & 0 & 0 \\
3 & 2 & 15 & 728 & 1 & 0  \\
4 & 3 & 63 & 6560 & 1 & 3 \\
5 & 4 & 255 & 59048 & 3 & 15 \\
6 & 5 & 1023 & 531440 & 17 & 216 \\
7 & 6 & 4095 & 4782968 & 34 & 1001 \\
8 & 7 & 16383 & 43046720 & 191 & 14168 \\
9 & 8 & 65535 & 387420488 & 500 & 77890 \\
10 & 9 & 262143 & 387420488 & 1886 & 652603 \\
11 & 10 & 1048575 & 3486784400 & 5981 & 5108147 \\
12 & 11 & 4194303 & 31381059608 & 30120 & 55906579 \\
13 & 12 & 16777215 & $2.54 \cdot 10^{12}$ & 68813 & 296956782 \\
14 & 13 & 67108863 & - & 429937 & - \\
15 & 14 & 268435455 & - & 1185306 & - \\
16 & 15 & 1073741823 & - & 4447563 & - \\
\hline
\end{tabular}
\label{tab:enum}
\end{table}
Remark that the column $\mathcal{M}_2$ reporting the numbers of maximum order invertible Sylvester matrices over $\F_2$ differs from the column $\#m\mathcal{LOCA}_d$ in Table II of our conference paper~\cite{mariot21}. Indeed, the latter is wrong, due to an error in how the order of the Sylvester matrix was computed in our previous Java implementation. Our new results in Table~\ref{tab:enum} have been double-checked by running the same algorithm in {\sc Magma} and computing the order of the matrix instead of checking the primitivity of the minimal polynomial, and we obtained the same results. Hence, we can be confident that the new counts reported in Table~\ref{tab:enum} are correct.

\section{Conclusions}
\label{sec:outro}
In this paper, we investigated a novel approach to generate pseudorandom sequences by means of cellular automata, namely by defining a dynamical systems based on two orthogonal CA. The trajectories of this system can be visualized as ``jumps'' over the superposed orthogonal Latin squares generated by the two CA, using the entries in each visited cell as the new set of row and column coordinates for the next cell. Remarking that two orthogonal CA induce a bijective superposition, the dynamics of the system is reversible and thus composed only of disjoint cycles. For this reason, we set up our investigation to search for OCA pairs that produce the largest cycles possible, which is a desirable property when considering pseudorandom generators for cryptographic applications. Further, the fact that the system is defined by a pair of orthogonal Latin squares implies that the update function is a multipermutation, which is a useful primitive when designing the diffusion layers of block ciphers.

We first performed an empirical search for the maximum cycle lengths of OCA pairs over the binary alphabet. This entailed first an exhaustive enumeration approach up to diameter $d=6$, and then an analysis of a sample of OCA pairs produced by the evolutionary algorithms described in~\cite{mariot17a} for $d=7$ and $d=8$. The results showed that there are both linear and nonlinear OCA pairs reaching the maximum possible cycle length of $2^{2n}-1$. Subsequently, we described a method to completely determine the cycle structure of linear OCA pairs, using the rational canonical form of the Sylvester matrix associated to the two linear rules. Further, observing that a Sylvester matrix has a maximum order of $q^{2n}-1$ if and only if its minimal polynomial is primitive, we devised an improved enumeration algorithm to generate them all for $q=2$ and $q=3$, respectively up to diameter $d=16$ and $d=13$. In doing that, we also fixed the numbers of such matrices for the binary case, which were reported incorrectly in the conference version of our paper~\cite{mariot21}.

There are several interesting directions and open problems for future research on this topic. The condition granted by Theorem~\ref{thm:prim} surely gives a better way to determine if a Sylvester matrix has maximum order than using Lagrange's theorem. However, it would be nice to give a precise characterization of when the minimal polynomial of a Sylvester matrix is primitive, which would probably yield a more efficient condition to check. The same goes also for a more precise characterization of the cycle sum of a Sylvester matrix. This would allow not only to determine the order of the matrix, but even to give a complete characterization of the cycles of a linear OCA pair. We are not aware of any result on the minimal polynomial or the rational canonical form of a Sylvester matrix, and we think this might be a good starting point for further research. Possibly, the minimal polynomial and rational canonical form in this case are related to the two polynomials that defines the Sylvester matrix.

A second interesting direction is to broaden the scope of the investigation to the more general nonlinear case. As we have seen in Section~\ref{subsec:exhaustive}, there exist also nonlinear OCA pairs that achieve a maximum cycle length of $2^{2n}-1$. One way to approach this problem would possibly be to consider the ANF of the local rules, and define a system of (multivariate) polynomial equations whose associated matrix resembles a Sylvester matrix, or one of its generalizations~\cite{gelfand08}. The study of nonlinear OCA pairs would also be interesting from a practical point of view. As a matter of fact, diffusion layers in block ciphers are usually implemented through linear transformations. Recently, however, there has been also an interest in \emph{nonlinear diffusion layers}~\cite{liu18}, which also provides a certain degree of confusion. Nonlinear OCA pairs could be considered for the design of such layers. More in general, one could also consider the use of nonlinear OCA pairs to design \emph{S-boxes}, which constitutes the confusion layer of block ciphers. There is quite an extensive body of literature concerning the design of S-boxes with good cryptographic properties based on CA, see for instance~\cite{szaban11,picek17,ghoshal18,mariot19}. Most of these works focus on the trade-off between reaching a high nonlinearity and a low differential uniformity to withstand certain attacks. In this respect, it would be interesting to determine whether the vectorial function $H$ defined by two nonlinear OCA pairs has also a good nonlinearity, and if the property of being a multipermutation positively affects the differential uniformity.

\section*{Appendix: Source Code and Experimental Data}
The source code of the algorithm and the experimental data discussed in this paper are available at \url{https://github.com/rymoah/hip-to-be-latin-square}.

\bibliographystyle{abbrv}
\bibliography{bibliography}

\end{document}